\documentclass[11pt]{article}

\usepackage{fancyhdr}

\usepackage{geometry}

\usepackage[dvips]{graphicx}

\usepackage{natbib}
\usepackage{geometry}

\renewcommand{\Re}{\mathbb{R}}

\newcommand{\vat}{\mathbb{E}}
\newcommand{\var}{\mathbb{V}\!\mathrm{ar}}

\newcommand\ind{\bot\hspace*{-6pt}\bot}  % {\bot\!\!\!\!\bot}
\newcommand{\bc}{\begin{center}}
	\newcommand{\ec}{\end{center}}

\newcommand{\bit}{\begin{itemize}}
	\newcommand{\eit}{\end{itemize}}

\newcommand{\be}{\begin{eqnarray*}}
	\newcommand{\ee}{\end{eqnarray*}}
\newcommand{\ben}{\begin{eqnarray}}
	\newcommand{\een}{\end{eqnarray}}
\newcommand{\g}{\,\vert\,}

\newcommand{\D}{\mathcal{D}}

\newcommand{\N}{\mathcal{N}}

\newcommand{\pa}{\mathrm{pa}}

\newcommand{\bzero}{\bm{0}}

\newcommand{\bD}{\bm{D}}

\newcommand{\bI}{\bm{I}}

\newcommand{\bL}{\bm{L}}

\newcommand{\bS}{\bm{S}}

\newcommand{\bU}{\bm{U}}
\newcommand{\bX}{\bm{X}}

\newcommand{\bZ}{\bm{Z}}

\newcommand{\bx}{\bm{x}}

\newcommand{\bz}{\bm{z}}

\newcommand{\bSigma}{\bm{\Sigma}}

\newcommand{\bOmega}{\bm{\Omega}}

\newcommand{\bmu}{\bm{\mu}}

\usepackage{tikz}
\usetikzlibrary{graphs}
\usetikzlibrary{arrows,automata,calc,shapes,positioning}

\definecolor{darkblue}{rgb}{0.0, 0.0, 0.65}

\newcommand{\black}{\color{black}}

\newcommand{\white}{\color{white}}

\usepackage{bm}
\usepackage{bbm}
\usepackage{float}
\usepackage{amssymb}
\usepackage{subfig}
\usepackage{enumerate}
\usepackage{longtable}
\usepackage{mathtools,cancel}
\usepackage{xurl}
\usepackage{multirow}
\usepackage{lscape}
\usepackage{authblk}

\begin{document}

	\title{Learning Bayesian networks: \\a copula approach for mixed-type data}
	\author[1]{Federico Castelletti \thanks{federico.castelletti@unicatt.it}}
	\affil[1]{Department of Statistical Sciences, Universit\`{a} Cattolica del Sacro Cuore, Milan}
	
	\date{}
	
	\maketitle
	
\begin{abstract}
	
	Estimating dependence relationships between variables is a crucial issue in many applied domains, such as medicine, social sciences and psychology.
	When several variables are entertained, these can be organized into a network which encodes their set of conditional dependence relations. Typically however, the underlying network structure is completely unknown or can be partially drawn only; accordingly it should be learned from the available data, a process known as \emph{structure learning}.
	In addition, data arising from social and psychological studies are often of different types, as they can include categorical, discrete and continuous measurements.
	In this paper we develop a novel Bayesian methodology for structure learning of directed networks which applies to 
	mixed data, i.e.~possibly containing continuous, discrete, ordinal and binary variables simultaneously.
	Whenever available, our method can easily incorporate known
	dependence structures among variables
	%partially-known directed dependencies between variables which can be postulated in advance based on available prior information.
	represented by paths or edge directions that can be postulated in advance based on the specific problem under consideration.
	We evaluate the proposed method through extensive simulation studies, with appreciable performances in comparison with current state-of-the-art alternative methods.
	Finally, we apply our methodology to well-being data
	%collected
	from a social survey
	promoted by the United Nations,
	and mental health data collected from a cohort of medical students.
	%R code implementing the proposed methodology is available at \url{https://github.com/FedeCastelletti/bayes_networks_mixed_data}.
	
	\vspace{0.7cm}
	
\end{abstract}

%%When you submit your PDF file or manuscript,
%%author names must be removed from all headers.
%%So if you use the \markright or \markboth commands,
%%and put your name in the header,
%%they must be disabled prior to your submission.
%%Before you can use either the \markright or \markboth
%%commands, you must disable the "\submit" switch above.
%\markright{}
%\markboth{}{}

%\author{Federico Castelletti}

%\affil{Department of Statistical Sciences
	%\\
	%Universit\`{a} Cattolica del Sacro Cuore, Milan}

%%\reprints{Correspondence should be sent to\\

%%\noindent E-Mail: federico.castelletti@unicatt.it \\
%\noindent Phone: \break
%\noindent Fax: \break
%%\noindent Website: https://sites.google.com/view/federicocastelletti}

\newpage

\section{Introduction}
\label{sec:introduction}

\subsection{Background and motivation}
\label{sec:introduction:motivation}

Learning dependence relations between variables is a pervasive issue in many applied domains, such as biology, social sciences, and notably psychology \citep{Briganti:Scutari:McNally:2022,Isvoranu:2022}.
In the latter context,
the recent
field of
\emph{network psychometrics} considers a network-based approach to represent psychological constructs and understand directed interactions between behavioral, cognitive and biological factors, possibly allowing for causal interpretations \citep{Borsboom:et:al:2021}.
The 2022 \textit{Psychometrika} special issue ``Network Psychometrics in action" promoted the development of statistical methods for network modelling motivated by psychological problems, and collected several contributions to the field, covering both methodological and applied aspects \citep{Marsman:Rhemtulla:2023:Editorial}.

Early works in the network psychometrics area were conceived to support psychologists in providing insights on various psychological phenomena, such as those at the basis of psychopatology, and in particular the study of comorbidity and mental disorders \citep{Borsboom:2008,Cramer:et:al:2010}.
Typical research questions thus relate to the identification of direct dependencies between manifest variables, differences in the underlying dependence structure across available groups of patients, or even to the design of clinical interventions based on an estimated network.
Crucial to these purposes is the development of models that allow to infer a plausible network structure for the available data and to provide a coherent quantification of the uncertainty related to directed links, specific paths or the whole network structure.
In this regard, methodologies that fully account for network uncertainty lead to parameter estimates that are more robust w.r.t. possible network-model misspecifications; see \citet{Haslbeck:and:Waldorp:2018}, \citet{Epskamp:et:al:2017} and \citet{Marsman:et:al:2022}
for recent contributions in this area inspired by psychological problems.

All of the issues introduced above have motivated the development of dedicated statistical methodologies, based both on a frequentist and on a Bayesian paradigm.
In particular, probabilistic \emph{graphical models} based on directed networks provide an effective tool to infer conditional dependence relations from the data \citep{Cowe:Dawi:Laur:Spie:1999,Edwa:2000}.
Additionally, Directed Acyclic Graphs (DAGs) offer a powerful framework for causal reasoning, even from observational, namely non-experimental, studies and specifically to quantify effects of hypothetical interventions on target variables w.r.t. outcome responses of interest; see \citet{Pear:2000} for a general introduction on causal inference based on DAGs, \citet{Maathuis:Nandy:Review} for a review.
The next section offers an overview of the main recent contributions to graphical modelling.
%that are related to the current work.

%Bayesian networks in the form of Directed Acyclic Graphs (DAGs) provide an effective tool to model complex dependence relations between variables in multivariate settings \citep{Cowe:Dawi:Laur:Spie:1999,Edwa:2000}.
%Applications of Bayesian networks are ubiquitous and especially in medicine, social sciences and psychology \citep{Frie:2004,Briganti:Scutari:McNally:2022}.
%There also exists a close connection between DAGs and Structural Equation Models (SEMs), which have become an established statistical technique
%in behavioral, educational and psychological studies to investigate conditional relations among variables of interest;
%see for instance \citet{Westland:2019} and \citet{McDonald:2002} for respectively a scholarly introduction and a more critical dissertation.
%Similarly to SEMs, DAGs adopt a graph structure to describe
%directional relationships between variables.
%In addition, they provide a powerful tool for causal inference, even from observational (i.e. non-experimental) studies \citep{Pear:2000}.
%In this framework, the notion of intervention and the allied do-calculus theory allow to quantify the causal effect on the mean level of a response of interest consequent to an intervention on any variable in the system; see also \citet{Maathuis:Nandy:Review} for a comprehensive review and \citet{Gische:Voelkle:2022}
%for extensions to broader classes of causal quantities that can be inferred from linearly parametrized graph-based models.

\subsection{Literature review}

%The network structure representing the set of dependence relations between variables is typically uncertain
%and accordingly must be inferred from the available data. In other cases, a graph structure can be partially drawn only whenever dependence relations between variables represented by directed paths can be postulated in advance. Examples are psychological studies where subject-specific features (e.g. sociological or demographic indicators) can be treated as exogenous variables, while other survey items as responses of interest possibly depending on the previous set of variables.
%
%medical science,
%where selected clinical features of patients are treated as exogenous variables, while a specific phenotype can be considered as a response of interest, and thus depending on the previous variables.
From a statistical perspective, learning a network of dependencies from the data is a model selection problem also known as \textit{structure learning}.
Several related methodologies that can deal with Gaussian and categorical data separately have been proposed.
Specifically, score-based
methods implement score functions for network estimation, such as based on
penalized maximum likelihood estimators \citep{Meinshausen:Buehlmann:2006, Frie:Etal:2008}, or marginal likelihoods for methodologies following a Bayesian perspective \citep{Hecketal:1995,Chic:2002}.
Moreover, constraint-based methods implement conditional independence tests to learn
the set of (in)dependence constraints characterizing the underlying DAG structure, as in the popular PC algorithm \citep{Spir:Glym:Sche:2000,Kalisch:Buehlmann:2007}.
%This starts from a complete, fully connected
%graph and then deletes recursively edges based on conditional independence decisions. The output is a Completed Partially Directed Acyclic Graph (CPDAG) representing the equivalence class of the estimated DAG.
%The original PC algorithm was introduced for both the Gaussian and categorical cases with suitable test statistics depending on the specific type of data. For instance, in the Gaussian setting conditional independence tests based on Pearson (partial) correlation coefficient are implemented.
On the other hand, Bayesian methodologies %for structure learning
adopt Markov chain Monte Carlo (MCMC) methods to approximate a posterior distribution over the space of network structures, or related features of interest; see for instance
\citet{Caste:etal:2018} and
\citet{Castelletti:Peluso:2021}
for respectively Gaussian and categorical settings, \citet{Ni:et:al:2022:Review} for a recent overview of Bayesian methods for structure learning with applications to biological problems.

Mixed-type data, i.e. observations from variables of different parametric families, are very common in
many contexts and expecially psychological studies, where ordinal, discrete and continuous measurements are simultaneously collected on subjects.
A few methodologies for structure learning from mixed data have been proposed.
\citet{Harris:Drton:2013:PC:paranormal} introduce the rank PC, an extension of the original PC algorithm
to nonparanormal models, namely based on a semi-parametric latent Gaussian copula model, with
purely continuous marginal distributions.
%Pearson correlations are replaced with rank-based measures such as Spearman's rank correlation and Kendall’s $\tau$ to estimate the dependence structure (correlation matrix) between observed nonparanormal variables.
Moreover,
\citet{Cui:et:al:2016:PC:copula} propose the Copula PC, an adaptation of the PC algorithm to a mixture of discrete 
and continuous data assumed to be drawn from a Gaussian
copula model. \citet{Cui:et:al:2018} extend the previous method to deal with data that are missing at random.
%Specifically, they apply rank correlations to pairwise complete observations and
%replace the sample size with an effective size in the conditional %independence tests to account for the information loss from missing values.
Similar ideas, for the case of undirected graphs, are also considered by
\citet{Muller:et:al:2019:copula:lasso}
and
\citet{He:et:al:2017:copula:graphical}.
%with directed graphs:
%\citet{Hobaek:Haff:et:al:2016}
Still in the context of directed graphs, a more recent methodology for structure learning given both categorical and Gaussian data is proposed by
%\citet{Bottcher:2001}
\citet{Andrews:et:al:2018}.
The authors introduce a mixed-variable polynomial score based on the notion of Conditional Gaussian (CG) distribution \citep{Lauritzen:Wermuth:1989}, then extended to a highly scalable algorithm by 
\citet{Andrews:et:al:PMLR:2019}.
Conditional Gaussian distributions are also adopted for structure learning of undirected graphs
by 
\citet{Lee:Hastie:PMLR:2013} and \citet{Cheng:et:al:2017} who implement penalized likelihoods and regression models with weighted lasso penalties respectively.
%using penalized likelihoods; second fit separate regressions with a weighted lasso penalty
In a Bayesian setting, \citet{Bhadra:et:al:20018} propose a unified framework for both categorical and Gaussian data based on Gaussian scale mixtures.

One main difficulty in developing statistical models for general mixed-type data is related to the non-standard joint support of the available variables.
Typically however, interest lies in estimating \emph{dependence parameters} of the joint distribution, corresponding to a network structure or correlation-type measures, rather than parameters indexing the marginal distributions of the variables.
In this context,
copula models, which allow to model the two sets of parameters separately, can provide an effective solution for statistical inference of network models.
In addition, \emph{semiparametric} copula models lacks any parametric assumption on the marginal c.d.f.'s which are estimated through their empirical distributions \citep{Hoff:2007}.
Contributions to copula graphical modelling based on undirected graphs are provided by \citet{Dobra:et:al:2011} and \citet{Mohammadi:Wit:2017}.

\subsection{Contribution and structure of the paper}

We propose a novel methodology for structure learning of networks which applies to mixed data, i.e.~comprising continuous, categorical as well as discrete and ordinal measurements.
Specifically, we consider a Gaussian copula model where the dependence parameter (covariance matrix) reflects the conditional independencies imposed by a directed acyclic graph, leading to our \emph{Gaussian copula DAG model}.
We consider a Bayesian framework and proceed by assigning suitable prior distributions to DAG structures and DAG-dependent parameters. Inference is carried out by implementing an MCMC scheme which approximates the posterior distribution over network structures and covariance matrices.
The main contributions of the proposed method can be summarized as follows:
i) we introduce a Bayesian framework for the analysis of complex dependence relations in multivariate settings characterized by mixed data;
ii) we provide a coherent quantification of the uncertainty around the estimated network or features of interest such as directed links, and a full posterior distribution of the underlying dependence parameter (correlation matrix), possibly summarized by Bayesian Model Averaging (BMA) estimates;
iii) our model allows to incorporate prior knowledge of the underlying network in terms of a partial ordering of the variables or edge orientations that are known in advance, thus improving DAG identification and enhancing causal inference.

The rest of the paper is organized as follows.
In Section 2
%\ref{sec:model}
we introduce Gaussian graphical models based on DAGs and the copula DAG model that we adopt for the analysis of mixed data.
Section 3
%\ref{sec:bayesian:inference}
completes our Bayesian model formulation by assigning prior distributions to DAG structures and DAG-model parameters.
We implement in Section 4
%\ref{sec:mcmc}
an MCMC scheme which approximates the posterior distribution of DAGs and parameters. Our method is evaluated through extensive simulation experiments in Section 5.
%\ref{sec:simulation:study}
%and compared with the benchmark Copula PC method.
Section 6
%\ref{sec:application}
is devoted to empirical studies, including the analysis of well-being data from a social survey promoted by the United Nations and mental health data collected from a cohort of medical students.
In Section 7
%\ref{sec:discussion}
we finally provide a discussion together with possible extensions of the proposed method to heterogeneous settings and latent trait models.
Additional simulation results, comparisons with alternative methods and examples of MCMC diagnostics of convergence are included in the Appendix.

\black

%%%%%%%%%%%%%%%%%%%
%%%%%%%%%%%%%%%%%%%

\section{Model specification}
\label{sec:model}

\subsection{Directed acyclic graphs}
\label{sec:dags}

A Directed Acyclic Graph (DAG) is a pair $\D=(V,E)$ consisting of a set of vertices (or nodes) $V=\{1,\dots,q\}$ and a set of directed edges $E \subseteq V \times V$.
For any two nodes $u,v \in V$, we denote an edge from $u$ to $v$ as $(u,v)$ or $u\rightarrow v$ indifferently; also, the set $E$ is such that if $(u,v)\in E$ then $(v,u)\notin E$.
A sequence of nodes $(v_1,v_2, \dots,,v_k)$ is a \emph{path} if there exists $v_1\rightarrow v_2\rightarrow \cdots\rightarrow v_k$ in $\D$. We assume that $\D$ does not contain \emph{cycles}, that is paths such that $v_1\equiv v_k$.
For a given node $v\in V$ we let $\pa_{\D}(v)$ be the set of parents of $v$ in $\D$, i.e. the set of all nodes $u$ such that $(u,v)\in E$.
%Moreover, we let $\fa(v) = v \cup \pa(v)$ be the \emph{family} of $v$ in $\D$.
%A DAG $\D$ is then \emph{complete} if all its nodes are connected.
Moreover, we say that $u$ is a \textit{descendant} of $v$ if there exists a path from $v$ to $u$; by converse, $v$ is an \textit{ancestor} of $u$. The set of all descendants and ancestors of a node $v$ in $\D$ are $\textnormal{de}_{\D}(v)$ and $\textnormal{an}_{\D}(v)$ respectively.

\begin{figure}
	\begin{center}
		\includegraphics[scale=1]{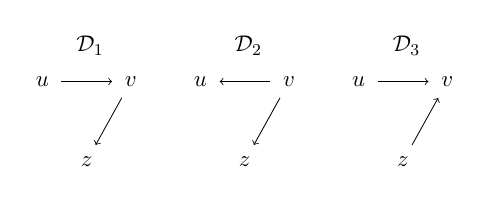}
	\end{center}
	\vspace{-0.5cm}
	\caption{Three DAGs on the set of nodes $V=\{u,v,z\}$. $\D_1$ and $\D_2$ encode the conditional independence $u \ind z \g v$. In $\D_3$ we instead have $u \ind z$.}
	\label{fig:dags}
\end{figure}

A DAG encodes a set of conditional independencies of the form $A\ind B \g C$, reading as \textit{``A and B are conditionally independent given $C$"}, where $A,B,C$ are disjoint subsets of the vertex set $V$.
The set of all conditional independencies characterizing the DAG determines the DAG \textit{Markov property} and
can be read-off from the graph using graphical criteria such as \textit{d-separation} \citep{Pear:2000}. In particular, each node is conditionally independent from its non descendants given its parents.
Simple examples are provided in Figure \ref{fig:dags}, where using d-separation it is possible to show that $u \ind z \g v$ in both $\D_1$ and $\D_2$; differently, $u \ind z$ in $\D_3$ meaning that $u$ and $z$ are marginally independent. \black We refer the reader to \citet{Laur:1996} for further notions on graph theory.

\subsection{Gaussian DAG models}
\label{sec:gaussian:dag}

Let $\D=(V,E)$ be a DAG and $\bz=(Z_1,\dots,Z_q)^{\top}$ a collection of $q$ real-valued random variables, each associated with a node in $\D$, and with joint p.d.f. $f(\cdot)$. 
We assume that
\ben
\label{eq:gaussian:DAG}
Z_1,\dots,Z_q \g \bOmega,\D \sim \N_q (\bzero, \bOmega^{-1}), \quad \bOmega \in \mathcal{P}_{\D},
\een
where $\bOmega$ is the precision matrix (inverse of the covariance matrix $\bSigma$) and $\mathcal{P}_{\D}$ denotes the set of all symmetric positive definite (s.p.d.) precision matrices Markov w.r.t. DAG $\D$.
Accordingly,
we impose to $\bOmega$ the conditional independencies encoded by $\D$ that are deducible from d-separation (Section 2.1). \black

An equivalent representation of Model \eqref{eq:gaussian:DAG}, useful for later developments, is given by the allied Structural Equation Model (SEM).
To this end, let $\bD=\textnormal{diag}(\bD_{11},\dots,\bD_{qq})$ and $\bL$ a $(q,q)$ matrix of (regression) coefficients with diagonal elements equal to 1 and $(u,v)$-element $\bL_{u,v}\ne 0$ if and only if $u\rightarrow v$ in $\D$.
Non-zero elements of $\bL$ correspond to directed links between nodes, while zero entries to missing edges in the DAG. Accordingly, $\bL$ resembles the DAG structure and the set of all parent-child relations characterizing its Markov property. \black
A Gaussian SEM can be written as
\ben
\label{eq:SEM}
\bL^{\top}\bz = \boldsymbol{\varepsilon}, \quad \boldsymbol{\varepsilon} \sim \N_q(\bzero,\bD),
\een
where $\bD$ corresponds to the covariance matrix of the error terms, which is assumed to be diagonal and collects the conditional variances of $(Z_1,\dots,Z_q)$.
\black
Moreover, Equation \eqref{eq:SEM}
implies $\var(\bz)=\bSigma=\bL^{-\top}\bD\bL^{-1}$; equivalently, $\bOmega=\bL\bD^{-1}\bL^{\top}$. The latter decomposition provides a re-parameterization of $\bOmega$ in terms of $(\bD,\bL)$. From \eqref{eq:SEM} we have, for each $j=1,\dots, q$,
%From this re-parameterization, the constraints imposed by $\D$ on the model parameters become more apparent since for each $(u,v)$-element of $\bL$, $u \ne v$, we have $\bL_{u,v}\ne 0$ if and only if $u \in \pa_{\D}(v)$, that is there is an edge $u\rightarrow v$ in $\D$. Hence, for $j=1,\dots,q$,
$
\label{eq:gaussian:linear:SEM}
Z_j = -\bL_{\prec j\,]}^{\top}\bz_{\pa_{\D}(j)} + \varepsilon_j,
$
with $\varepsilon_j\sim\N(0,\bD_{jj})$,
where $\prec j\,]\,=\pa_{\D}(j)\times j$ and $\bL_{A\times B}$ denotes the sub-matrix of $\bL$ with elements belonging to rows and columns indexed by $A$ and $B$ respectively.
Each equation above resembles the structure of a linear regression model for variable $Z_j$, with $-\bL_{\prec j\,]}$ corresponding to the regression coefficients associated with variables in $\bz_{\pa_{\D}(j)}$, namely the parents of node/variable $Z_j$; see also Section 3.2. for a comparison with the Bayesian analysis of normal linear regression models.
\black
Accordingly, Model \eqref{eq:gaussian:DAG} can be equivalently written as
\ben
\label{eq:gaussian:DAG:model}
f(z_1,\dots,z_q\g\bD,\bL,\D)=\prod_{j=1}^{q}\,d\,\N\big(z_j\g -\bL_{\prec j\,]}^{\top}\bz_{\pa_{\D}(j)}, \bD_{jj}\big),
\een
where $d\,\N(\cdot\g \mu,\sigma^2)$ denotes the p.d.f.~of a univariate $\N(\mu,\sigma^2)$.
Finally, given $n$ i.i.d.~samples from \eqref{eq:gaussian:DAG:model}, $\bz_i=(z_{i,1},\dots,z_{i,q})^{\top}$, $i=1,\dots,n$, collected in the $(n,q)$ matrix $\bZ$ (row-binding of the $\bz_i$'s),
the likelihood function can be written as
\ben
\label{eq:like:D:L}
f(\bZ \g \bD,\bL, \D)=
\prod_{i=1}^{n}\left\{
\prod_{j=1}^{q}d\,\N\big(z_{i,j}\g -\bL_{\prec j\,]}^{\top}\bz_{i,\pa_{\D}(j)}, \bD_{jj}\big)\right\}.
\een

\subsection{Copula DAG models}
\label{sec:gaussian:copula}

Consider now a collection of $q$ random variables, $X_1,\dots,X_q$, comprising binary, ordinal, continuous or count variables, each with marginal cumulative distribution function (c.d.f.)~$F_j(\cdot)$, $j=1,\dots,q$.
In what follows we will consider a collection of $n$ $q$-dimensional observations from $X_1,\dots,X_q$.
To model these \emph{mixed} data we need to specify a joint distribution for $X_1,\dots, X_q$ that we specify through a 
Gaussian \emph{copula} DAG model.
%In doing this, a first complication arise since the marginal c.d.f.'s belong to different (possibly non-standard) parametric families.
%However, our general interest lies in the estimation of the \emph{association parameters} rather than the parameters of the marginal c.d.f.'s.
%A \emph{copula model} allows to model separately the association parameters from the marginal distributions and therefore provides an effective solution to this issue.
%Moreover, a \emph{semiparametric} copula models lacks any parametric assumption on $F_1,\dots,F_q$ which are estimated through their empirical distributions \citep{Hoff:2007}.
Specifically, let $Z_1,\dots,Z_q$ be a collection of $q$ latent random variables with joint Gaussian distribution as in \eqref{eq:gaussian:DAG},
%that is
%\ben
%Z_1,\dots,Z_q \g \bOmega,\D \sim \N_q (\bzero, \bOmega^{-1}), \quad \bOmega \in \mathcal{P}_{\D},
%\een
%where again $\bOmega$ is the precision matrix (inverse of the covariance matrix $\bSigma$) and $\mathcal{P}_{\D}$ is set of all s.p.d. precision matrices Markov w.r.t. DAG $\D$.
We establish a link between each observed variable $X_j$ and its latent counterpart $Z_j$ by assuming that
\ben
\label{eq:link:latent:observed}
X_j = F_j^{-1}\left\{\Phi(Z_j)\right\},
\een
% where $\Phi(\cdot)$ is the c.d.f. of a standard normal distribution
where $F^{-1}_j$ is the (pseudo) inverse c.d.f.~of $X_j$ and $\Phi(Z_j)$ the c.d.f.~of a standard Normal distribution.
%We remark that the adoption of a \textit{pseudo} inverse is justified by the presence of variables whose marginal distributions are not continuous.
The joint c.d.f.~of $X_1,\dots,X_q$ can be written as
\ben
\label{eq:copula:cdf}
\begin{aligned}
P(X_1\le x_1,\dots, X_q \le x_q\g \bOmega,F_1,\dots,F_q) \hspace{3cm}\\=
\Phi_q\left(\Phi^{-1}(F_1(x_1)),\dots,\Phi^{-1}(F_q(x_q)) \g \bOmega\right),
\end{aligned}
\een
where $\Phi_q(\cdot \g \bOmega)$ denotes the c.d.f.~of $\N_q(\bzero,\bOmega^{-1})$ in \eqref{eq:gaussian:DAG}.
Also notice that Model \eqref{eq:copula:cdf} depends on the marginal distributions $F_1,\dots,F_q$ and (although not emphasized in the equation) their parameters which would need to be ``estimated".
A semiparametric estimation strategy
%suggested by \citet{Hoff:2007} and that we adopt throughout this work,
would replace $F_j$ with the corresponding empirical estimates
$
\widehat{F}_j(k_j)=n^{-1}\sum_{i=1}^{n}\mathbbm{1}(x_{i,j}<k_j),
$
where $k_j\in \textnormal{unique}\{x_{1,j},\dots,x_{n,j}\}$.
A comparison with a parametric strategy based on probabilistic-model assumptions for the marginal c.d.f.'s is instead provided in the Appendix. \black

As an alternative to the estimation procedures above, \citet{Hoff:2007} proposes a \emph{rank-based} non-parametric approach, that we also employ in our methodology.
Specifically, let $\bx_i=(x_{i,1},\dots,x_{i,q})^{\top}$, $i=1,\dots,n$, be $n$ i.i.d.~samples from \eqref{eq:copula:cdf} and $\bX$ the $(n,q)$ data matrix.
Since the $F_j$'s are non decreasing, for each pair of distinct observations $x_{i,j}$ and $x_{l,j}$, if $x_{i,j}<x_{l,j}$ then $z_{i,j}<z_{l,j}$. Therefore, observing $\bX$ implies that the latent data $\bZ$ must lie in the set
\ben
\label{eq:set:A}
A(\bX) = \left\{\bZ \in \Re^{n\times q} : \textnormal{max}\left\{z_{k,j}:x_{k,j}<x_{i,j}\right\} <
z_{i,j} < \textnormal{max}\left\{z_{k,j}:x_{i,j}<x_{k,j} \right\}\right\} \nonumber \\
\een
and one can take the occurrence of such an event as the data.
Thus, the \emph{extended rank likelihood} \citep{Hoff:2007} can be written as
\ben
\label{eq:rank:likelihood}
p(\bZ\in A(\bX)\g \bD, \bL, \D)=\int_{A(\bX)}f(\bZ\g\bD, \bL, \D)\, d\bZ,
\een
with $f(\bZ\g\bD, \bL, \D)$ as in Equation \eqref{eq:like:D:L}.

Our model formulation assumes that a DAG Markov property holds at a latent-variable stage, namely between $Z_1,\dots,Z_q$, in force of factorization \eqref{eq:gaussian:DAG:model}.
Through the copula-transfer link \eqref{eq:link:latent:observed}, this translates into a Markov property for $X_1,\dots,X_q$, provided that all the marginal distributions are continuous \citep{Liu:et:al:2009}.
The presence of non-continuous variables (e.g.~ordinal, discrete) might induce additional dependencies among the observed variables (w.r.t.~those included in the latent model); however such dependencies can be regarded as having a secondary relevance since they emerge from the marginals, rather than from the joint distribution \citep{Dobra:et:al:2011}.

\black

\section{Bayesian inference}
\label{sec:bayesian:inference}
We complete the Gaussian copula DAG model introduced in the previous section by assigning prior distributions to parameters $(\bD,\bL,\D)$.
Since parameters $(\bD,\bL)$ are DAG-dependent, as they satisfy structural constraints imposed by the DAG (Section 2.2) we structure our prior as $p(\bD,\bL,\D)=p(\bD,\bL\g\D)p(\D)$.

\subsection{Prior on $\D$}
\label{sec:prior:DAG}

Let $\mathcal{S}_q$ be the set of \emph{all} DAG structures on $q$ nodes.
In many applied problems,
exact knowledge about the orientation of some edges, whenever present in the graph, is typically available and accordingly one would like to incorporate such information in the model.
Without loss of generality, let $\mathcal{S}_q^C$ be the set of all DAGs on $q$ nodes satisfying a set of structural constraints $C$, corresponding to edge orientations that are known in advance (equivalently, a set of ``reversed" edge orientations that are forbidden). \black
%and introduce a \textit{partial ordering} of the nodes in the graph.
As an example, suppose node $u$ represents a response variable of interest, so that there are no outgoing edges \emph{from} $u$ (equivalently, $u$ cannot have children). In such a case we have $C=\{u\not\rightarrow v \g v=1,\dots,q, v \ne u\}$ and accordingly an edge between $u$ and $v$ whenever present will be oriented as $v\rightarrow u$.
%in other words, $u$ cannot have \textit{children}.
See also Section %\ref{sec:application}
6
for examples on real-data.
We assign a prior to DAGs belonging to $\mathcal{S}_q^C$ as follows.

%, through a Beta-Binomial distribution on the number of edges in the graph. 
For a given DAG $\D=(V,E)\in\mathcal{S}_q^C$, let $\bS^{\D}$ be the $0-1$ \emph{adjacency matrix} of its skeleton, that is the underlying undirected graph obtained after removing the orientation of all its edges. For each $(u,v)$-element of $\bS^{\D}$, we have $\bS_{u,v}^{\D}=1$ if and only if $(u,v)\in E$ or $(v,u)\in E$, zero otherwise.
Conditionally on a prior probability of inclusion $\pi\in(0,1)$ we assume for each $u>v$,
$
\bS_{u,v}^{\D} \g \pi \overset{\textnormal{iid}}\sim \text{Ber}(\pi),
$
which implies
\ben
p(\bS^{\D}\g \pi)=\pi^{|\bS^{\D}|} (1-\pi)^{\frac{q(q-1)}{2}-|\bS^{\D}|},
\een
where $|\bS^{\D}|$ is the number of edges in $\D$ (equivalently in its skeleton) and $q(q-1)/2$ is the maximum number of edges in a DAG on $q$ nodes.
We then assume $\pi\sim \textnormal{Beta}(c,d)$, so that, by integrating out $\pi$, the resulting prior on $\bS^{\D}$ is
\ben
\label{eq:prior:multiplicity}
p(\bS^{\D}) = \frac
{\Gamma \left(|\bS^{\D}| + c\right)\Gamma \left(\frac{q(q-1)}{2} - |\bS^{\D}| + d \right)}
{\Gamma \left(\frac{q(q-1)}{2} + c + d\right)}
\cdot
\frac
{\Gamma \left(c + d\right)}
{\Gamma \left(c\right)\Gamma \left(d\right)}.
\een
%A similar prior was introduced by \citet{Scott:Berger:2010} for variable selection in linear models, where it was also shown to account for multiplicity correction.
Finally, we set
$
%\label{eq:prior:DAG}
p(\D)\propto p(\bS^{\D})$ for each $\D\in\mathcal{S}_q^C
$.
See also \citet{Scot:Berg:2010} for a comparison with multiplicity correction priors adopted in a linear model selection setting.
Hyperparameters $c$ and $d$ can be chosen to reflect a prior knowledge of sparsity in the graph, whenever available; in particular any choice $c < d$ will imply $\vat(\pi)<0.5$, thus favoring sparse graphs.
The default choice $c=d=1$, which corresponds to $\pi \sim \textnormal{Unif}(0,1)$, can be instead adopted in the absence of substantive prior information.
\black

\subsection{Prior on $(\bD,\bL)$}

Conditionally on DAG $\D$ we assign a prior to $(\bD,\bL)$
through a DAG-Wishart distribution with 
position hyperparameter $\bU$ (a $(q,q)$ s.p.d.~matrix)
and shape hyperparameter $a^{\D}=(a_1^{\D},\dots,a_q^{\D})^{\top}$.
The DAG-Wishart distribution \citep{cao:et:al:2019} provides a conjugate prior for the Gaussian DAG model \eqref{eq:gaussian:DAG:model}. Accordingly, conditionally on the latent data $\bZ$, the posterior distribution of $(\bD,\bL)$ as well as the marginal likelihood of the model are available in closed form; see Section 4 for more details.
A feature of the DAG-Wishart distribution is also that node-parameters $\big\{
(\bD_{jj}, \bL_{\prec j\,]} ), \,
j=1,\dots,q
\big\}$ are \textit{a priori} independent with distribution
\ben
\label{eq:prior:cholesky}
\begin{aligned}
	\bD_{jj} \g \D \sim& \,\,\, \textnormal{I-Ga}\left(\frac{1}{2}a_j^{\D},
	\frac{1}{2}\bU_{j\g\pa_{\D}(j)}\right), \\
	\bL_{\prec j\,]}\g\bD_{jj}, \D \sim& \,\,\, \N_{|\pa_{\D}(j)|}\left(-\bU_{\prec j \succ}^{-1}\bU_{\prec j\,]},\bD_{jj}\bU_{\prec j \succ}^{-1}\right),
\end{aligned}
\een
where $\textnormal{I-Ga}(\alpha,\beta)$ stands for an Inverse-Gamma distribution with shape $\alpha>0$ and rate $\beta>0$ having expectation $\beta/(\alpha-1)$ ($\alpha>1$).
Moreover,
$\bU_{j\g\pa_{\D}(j)} = \bU_{jj}-\bU_{[\, j\succ} \bU^{-1}_{\prec j\succ} \bU_{\prec j\,]}$, with
$\prec j\,] = \pa_{\D}(j)\times j$, $[\, j\succ\, = j \times \pa_{\D}(j)$, $\prec j\succ\, = \pa_{\D}(j)\times \pa_{\D}(j)$.
\black
With regard to hyperparameters $a_1^{\D},\dots,a_q^{\D}$ we also consider the default choice
$a_j^{\D}=a+|\pa_{\D}(j)|-q+1$ $(a>q-1)$ which
guarantees \textit{compatibility} (same marginal likelihood) among prior distributions for Markov equivalent DAGs; see also \cite{Peluso:Consonni:2020}.
Finally, the prior on $(\bD,\bL)$ is given by
\ben
\label{eq:prior:indep:chol}
p(\bD,\bL\g\D)=\prod_{j=1}^{q}p(\bL_{\prec j\,]}\g\bD_{jj})\,p(\bD_{jj}).
\een

A DAG-Wishart prior on $(\bD,\bL)$ implicitly assigns (independent) Normal-Inverse-Gamma distributions to each pair of node-parameters $(\bD_{jj},\bL_{ \prec j\,]})$, a conditional variance and vector-regression coefficient for the $j$-th term of the SEM, as in the standard conjugate Bayesian analysis of a normal linear regression model.
Moreover, under the default choice $\bU=g\bI_q$ (Section 5), with $g>0$ and $\bI_q$ the $(q,q)$ identity matrix, it is easy to show that $\vat(\bL_{ \prec j\,]}\g \bD_{jj}, \D)=\bzero$ and $\var(\bL_{ \prec j\,]}\g \bD_{jj}, \D)=\bD_{jj}/g\bI_{|\pa_{\D}(j)|}$ for each $j=1,\dots,q$, so that priors on regression coefficients are centered at zero and with diagonal covariance matrix reflecting an assumption of prior independence across elements of $\bL_{ \prec j\,]}$; moreover, smaller values of $g$ make such prior less informative; we refer to Section 5 for details about the choice of hyperparameters.
\black

\section{Computational implementation and posterior inference}
\label{sec:mcmc}

Our target is the joint posterior of $(\bD,\bL,\D,\bZ)$, namely
\ben
\label{eq:posterior}
p(\bD,\bL,\D,\bZ\g \bX)\propto p(\bZ\in A(\bX)\g\bD,\bL,\D)p(\bD,\bL\g\D)p(\D),
\een
where $p(\bZ\in A(\bX)\g\bD,\bL,\D)$ is the extended rank likelihood in \eqref{eq:rank:likelihood}, $p(\D)$ and $p(\bD,\bL\g\D)$ the priors on DAG $\D$ and DAG-parameters $(\bD,\bL)$ introduced in Section 3 %\ref{sec:bayesian:inference}
respectively.
An MCMC scheme targeting the posterior \eqref{eq:posterior} can be constructed by iteratively sampling $\D$, $(\bD,\bL)$ and $\bZ$ from their full conditional distributions as we detail in the following.

\subsection{4.1 Update of $(\bD,\bL,\D)$}
\label{sec:mcmc:full:cond:DL:dag}

The joint full conditional distribution of $(\bD,\bL,\D)$ is given by
\ben
p(\bD,\bL,\D\g\bX,\bZ) = p(\bD,\bL,\D\g\bZ) \propto f(\bZ\g\bD,\bL,\D)p(\bD,\bL\g\D)p(\D). \quad\quad
\een
Conditionally on the latent data $\bZ$, we can sample from $p(\bD,\bL,\D\g\bZ)$ using the MCMC scheme for posterior inference of Gaussian DAGs presented in \citet[Supplementary Material]{Castelletti:Consonni:2021:BA}. The latter consists of a Partial Analytic Structure (PAS) algorithm \citep{Godsill:2012} based on the two following steps.

\vspace{0.3cm}

Given DAG $\D$, parameters $(\bD,\bL)$ are first sampled from their full conditional distribution, which, because of conjugacy of the DAG-Wishart prior with the distribution of the latent data, is such that, for $j=1,\dots,q$,
\ben
\label{eq:posterior:D:L}
\begin{aligned}
	\bD_{jj} \g \D, \bZ \sim& \,\,\, \textnormal{I-Ga}\left( \frac{1}{2}\widetilde a_j^{\D},
	\frac{1}{2}\widetilde\bU_{j\g\pa_{\D}(j)}\right), \\
	\bL_{\prec j\,]}\g\bD_{jj}, \D, \bZ \sim& \,\,\, \N_{|\pa_{\D}(j)|}\left(-\widetilde\bU_{\prec j \succ}^{-1}\widetilde\bU_{\prec j\,]},\bD_{jj}\widetilde\bU_{\prec j \succ}^{-1}\right),
\end{aligned}
\een
with
$\widetilde a_j^{\D} = \widetilde{a}+|\pa_{\D}(j)|-q+1$, $\widetilde{a} = a+n$ and
$\widetilde \bU = \, \bU + \bZ^\top\bZ$.

\vspace{0.4cm}

Next, update of DAG $\D$ is performed through a Metropolis Hastings step in which a DAG $\D^*$ is drawn from a proposal distribution $q(\D^*\g\D)$.
For a given DAG $\D\in\mathcal{S}_q^C$, the adopted proposal distribution is built upon the set of all DAGs belonging to $\mathcal{S}_q^C$ that can be reached from $\D$ through insertion, deletion or reversal of an edge in $\D$.
Specifically, we construct the set of all \textit{direct successors} of $\D$, $\mathcal{O}_{\D}$, and then draw uniformly a DAG $\D^*$ from $\mathcal{O}_{\D}$. It follows that $q(\D^*\g\D)=1/|\mathcal{O}_{\D}|$.
%In the first step, for a given DAG $\D$, a new DAG $\widetilde{\D}$ is proposed from a suitable proposal distribution which is defined as follows.
%We consider three types of operators that locally modify a DAG: insert a directed edge (InsertD $u \rightarrow v$
%for short), delete a directed edge (DeleteD $u\rightarrow v$) and reverse a directed edge (ReverseD
%$u \rightarrow v$). For a given $\D \in \mathcal{S}_q$, where $\mathcal{S}_q$ is the set of all DAGs on $q$ nodes, we  construct the set of valid operators $\mathcal{O}_{\D}$, that is operators whose resulting graph is a DAG.
%A DAG $\widetilde{\D}$ is then called a \emph{direct successor} of $\D$ if it can be reached by applying an operator in $\mathcal{O}_{\D}$ to $\D$.
%Therefore, given the current $\D$ we propose $\widetilde{\D}$ by uniformly sampling an element in $\mathcal{O}_{\D}$ and applying it to $\D$. Since there is a one-to-one correspondence between each operator and the resulting DAG, the probability of transition is $q(\widetilde{\D}\g\D)=1/|\mathcal{O}_{\D}|$, for each $\widetilde{\D}$ direct successor of $\D$.
Each proposed DAG then differs \emph{locally} by a single edge $(u,v)$ which is inserted $(a)$, deleted $(b)$ or reversed $(c)$ in $\D$.
It can be shown that
the acceptance probability for $\D^*$ under a PAS algorithm is given by
$\alpha_{\D^*}=\min\{1;r_{\D^*}\}$
with
\ben
r_{\D^*}\,=\,
\left\{
\begin{array}{rl}
	\dfrac{m_{}(\bZ_v\g\bZ_{\pa_{\D^*}(v)}, \D^*)}
	{m_{}(\bZ_v\g\bZ_{\pa_{\D}(v)}, \D)}
	\cdot\dfrac{p(\D^*)}{p(\D)}
	\cdot\dfrac{q(\D\g\D^*)}{q(\D^*\g\D)} & \quad \\ \textnormal{ if } (a) \textnormal{ or } (b) \vspace{0.2cm} \\
	\dfrac{m_{}(\bZ_v\g\bZ_{\pa_{\D^*}(v)}, \D^*)}
	{m_{}(\bZ_v\g\bZ_{\pa_{\D}(v)}, \D)} \cdot
	\dfrac{m_{}(\bZ_u\g\bZ_{\pa_{\D^*}(u)}, \D^*)}
	{m_{}(\bZ_u\g\bZ_{\pa_{\D}(u)}, \D)}
	\cdot\dfrac{p(\D^*)}{p(\D)}
	\cdot\dfrac{q(\D\g\D^*)}{q(\D^*\g\D)} & \quad \\ \textnormal{ if } (c),
\end{array}
\white\right\}
\een
where
\be
m\left(\bZ_v\g\bZ_{\pa_{\D}(v)}, \D\right) =
(2\pi)^{-\frac{n}{2}}\cdot
\frac{\big|\bU_{\prec u \succ}\big|^{\frac{1}{2}}}
{\big|\widetilde{\bU}_{\prec v \succ}\big|^{\frac{1}{2}}}\cdot
\frac{\Gamma\left(\frac{1}{2}\widetilde{a}_v^{\D}\right)}
{\Gamma\left(\frac{1}{2}a_v^{\D}\right)}\cdot
\frac{\Big(\frac{1}{2}\bU_{v\g\pa_{\D}(v)}\Big)^{\frac{1}{2}a_v^{\D}}}
{\left(\frac{1}{2}\widetilde{\bU}_{v\g\pa_{\D}(v)}\right)^{\frac{1}{2}\widetilde{a}_v^{\D}}}
\ee
is the marginal (i.e. integrated w.r.t.~$\bL_{\prec v\,]}$ and $\bD_{vv}$) data distribution relative to the conditional density
of $Z_v$ given $\bz_{\pa_{\D}(v)}$ appearing in \eqref{eq:like:D:L}. 
We refer the reader to \citet{Castelletti:Mascaro:2022:BCDAG} for further computational details.

\subsection{Update of $\bZ$}
\label{sec:mcmc:full:cond:Z}

Since the latent data $\bZ$ are known only relative to the event $A(\bX)$ (Equation \eqref{eq:set:A}), we also need to sample them from their full conditional distribution. The latter is given by
\ben
p(\bZ\g\bD,\bL,\D,\bX) \propto p(\bZ \in A(\bX) \g\bD,\bL,\D).
\een
Also, by exploiting the structure of the extended rank likelihood and the set $A(\bX)$ in
\eqref{eq:rank:likelihood} and \eqref{eq:set:A} respectively,
the conditional distribution of $Z_{i,j}$ corresponds to a 
$\N(z_{i,j}\g -\bL_{ \prec j\,]}^\top\bz_{i,\pa(j)}, \bD_{jj})$ truncated at $\big(z_{i,j}^{(l)}, z_{i,j}^{(u)}\big)$, where $z_{i,j}^{(l)}=\textnormal{max}\{z_{k,j}:x_{k,j}<x_{i,j}\}$ and
$z_{i,j}^{(u)}=\textnormal{max}\{z_{k,j}:x_{i,j}<x_{k,j} \}$.
Therefore, conditionally of the current values of $\bD$ and $\bL$, update of the $(n,q)$ data matrix $\bZ$ can be performed by drawing each latent observation $z_{i,j}$, $j=1,\dots,q$, $i=1,\dots,n$, from its corresponding truncated-Normal conditional distribution.

\subsection{Posterior inference}

The output of our MCMC scheme is a collection of DAGs $\big\{\D^{(s)}\big\}_{s=1}^S$ and DAG-parameters $\big\{\big(\bD^{(s)}, \bL^{(s)}\big)\big\}_{s=1}^S$
approximately sampled from \eqref{eq:posterior}, where $S$ is the number of fixed MCMC iterations.
An approximate posterior distribution over the space of DAGs can be obtained by computing, for each $\D\in\mathcal{S}_q^C$,
\ben
\widehat{p}(\D\g \bX)
=\frac{1}{S}
\sum_{s=1}^{S}\mathbbm{1}\left\{\D^{(s)}=\D\right\},
\een
which approximates the posterior probability of each DAG structure through its MCMC frequency of visits.
As a summary of the previous output, we can also recover a $(q,q)$ matrix of (marginal) posterior probabilities of edge inclusion, whose $(u,v)$-element corresponds to
\ben
\label{eq:posterior:probability:edge:inclusion}
\widehat p (u\rightarrow v \g \bX)
\equiv \widehat p_{u\rightarrow v}
=\frac{1}{S}\sum_{s=1}^S\mathbbm{1}_{u\rightarrow v}\big\{\D^{(s)}\big\},
\een
an MCMC frequency-based estimated posterior probability of $u \rightarrow v$, where $\mathbbm{1}_{u\rightarrow v}\big\{\D^{(s)}\big\}=1$ if $\D^{(s)}$ contains $u\rightarrow v$, $0$ otherwise.
From the previous quantities, a single graph estimate summarizing the entire MCMC output can be also recovered. Specifically, by fixing a threshold for edge inclusion $k \in (0,1)$, a DAG estimate can be obtained by including those edges whose posterior probability of inclusion in \eqref{eq:posterior:probability:edge:inclusion} exceeds $k$. When $k=0.5$, we name the resulting estimate \emph{Median Probability DAG Model} (MPM DAG), following the idea proposed by \citet{Barb:Berg:2004} in a linear regression context\black. 
Finally, we can recover a posterior summary of the covariance matrix $\bSigma$ through the corresponding Bayesian Model Averaging \citep[BMA]{Hoeting:et:al:1999:bma} estimate 
\ben
\label{eq:BMA:covariance}
\widehat{\bSigma}
=\frac{1}{S}
\sum_{s=1}^{S}\bSigma^{(s)},
\een
where for each $s=1,\dots,S$, $\bSigma^{(s)}=\big(\bL^{(s)}\big)^{-\top}\bD^{(s)}\big(\bL^{(s)}\big)^{-1}$.

\section{Simulation study}
\label{sec:simulation:study}

In this section we evaluate the performance of the proposed method through simulations.

\subsection{Scenarios}
\label{sec:simulation:study:scenarios}

We fix the number of variables $q=20$ and consider different simulation settings where data are generated by varying the following features:
\begin{itemize}
\item[] $(a)$ the class of DAG structure;
\item[] $(b)$ the type of variables;
\item[] $(c)$ the sample size.
\end{itemize}
With regard to $(a)$, we consider three different classes of DAG structures:
\begin{enumerate}
\item[] $(a.1)$ \textit{Free}: no structural constraints imposed to the DAG;
\item[] $(a.2)$ \textit{Regression}: we fix each node $u\in\{1,2,3\}$ as a response, so that edges $u\rightarrow v$, for $v \in V \setminus \{1,2,3\}$ are not allowed;
\item[] $(a.3)$ \textit{Block}: nodes are partitioned into two sets of equal size $A$ and $B$ and only edges within each set or from set $A$ to $B$ are allowed.
\end{enumerate}
Under each scenario defined by $(a)$, $N=40$ DAGs are generated by fixing a probability of edge inclusion $\pi=0.1$ for all edges which are allowed in the corresponding class of DAG structures $(a)$.
For each given DAG $\D$, we generate the parameters of the underlying Gaussian DAG model in \eqref{eq:gaussian:DAG:model} by fixing $\bD = \bI_q$, while drawing non-zero elements of $\bL$ uniformly in the interval $[-1,-0.1] \cup [0.1,1]$.
Latent observations collected in the $(n,q)$ data matrix $\bZ$ are then generated according to \eqref{eq:gaussian:DAG:model} for different sample sizes $n \in \{100,200,500,1000,2000\}$ $(c)$.
Starting from the latent data matrix $\bZ$, observed data are finally generated as in 
\eqref{eq:link:latent:observed} for different choices of the marginal c.d.f's $F_1,\dots,F_q$ $(b)$. Specifically, we consider the following assumptions:
\begin{enumerate}
\item[] $(b.1)$ \textit{Binary}: $X_j\sim\textnormal{Ber}(\eta_j)$, $j=1,\dots,q$, with $\eta_j$ randomly drawn in $[0.2,0.8]$;
\item[] $(b.2)$ \textit{Ordinal}: $X_j\sim\textnormal{Binomial}(\theta_j,5)$, $j=1,\dots,q$, with $\theta_j$ randomly drawn in $[0.2,0.8]$;
\item[] $(b.3)$ \textit{Count}: $X_j\sim\textnormal{Poisson}(\lambda_j)$, $j=1,\dots,q$, with $\lambda_j$ randomly drawn in $[1,10]$;
\item[] $(b.4)$ \textit{Mixed}: $X_1\dots,X_7$ as in \textit{Binary}; $X_8\dots,X_{15}$ as in \textit{Ordinal};  $X_{16}\dots,X_{20}$ as in \textit{Count}.
\end{enumerate}
Each combination of $(a)$, $(b)$ and $(c)$ defines a \textit{simulation scenario} which consists of $N=40$ (true) DAGs and allied datasets.

\subsection{Results}

We run our MCMC scheme to approximate the posterior distribution of DAG structures and parameters.
In particular, under each scenario among \textit{Free, Regression, Block}, we limit the DAG space $\mathcal{S}_q^C$ to those DAGs satisfying the constraints imposed by the corresponding class of DAG structures; see also Section 3.
The number of iterations was fixed as $S=15000$ after some pilot simulations that were used to assess the convergence of the MCMC chain.
We also set $\bU=g\bI_q$ with $g = 1/n$, $a = q$ in the DAG-Wishart prior \eqref{eq:prior:cholesky}, while $c = 1$, $d = 5$ in \eqref{eq:prior:multiplicity}, which corresponds to a $\textnormal{Beta}(1,5)$ prior on the probability of edge inclusion $\pi$ and implies $\vat(\pi)=0.1\bar{6}$. While this specific choice of $c$ and $d$ can favour \textit{sparsity} in the graphs, some further analyses not reported for brevity showed that results are quite insensitive to the values of the two hyperparameters.

\vspace{0.4cm}

\begin{landscape}
\begin{figure}
	\begin{center}
		\begin{tabular}{lc}
			%\textit{Sparse Scenario} \quad\quad
			%\\
			\multirow{1}{*}{\rotatebox[origin=c]{90}{\textit{Free \quad\, }}} & \raisebox{-0.7\height}{\includegraphics[scale=0.50]{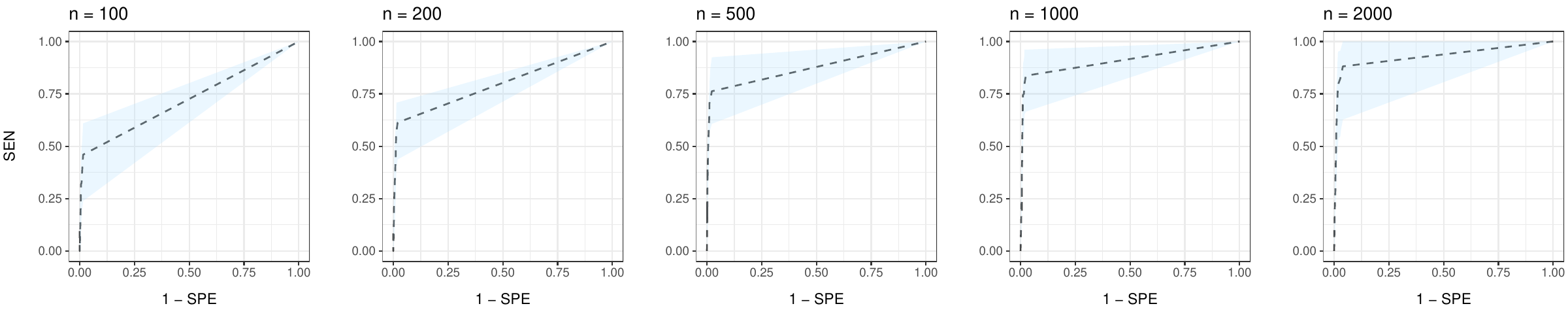}}
			\vspace{0.4cm}
			\\
			\multirow{1}{*}{\rotatebox[origin=c]{90}{\textit{Regression\, }}} &
			\raisebox{-0.7\height}{\includegraphics[scale=0.50]{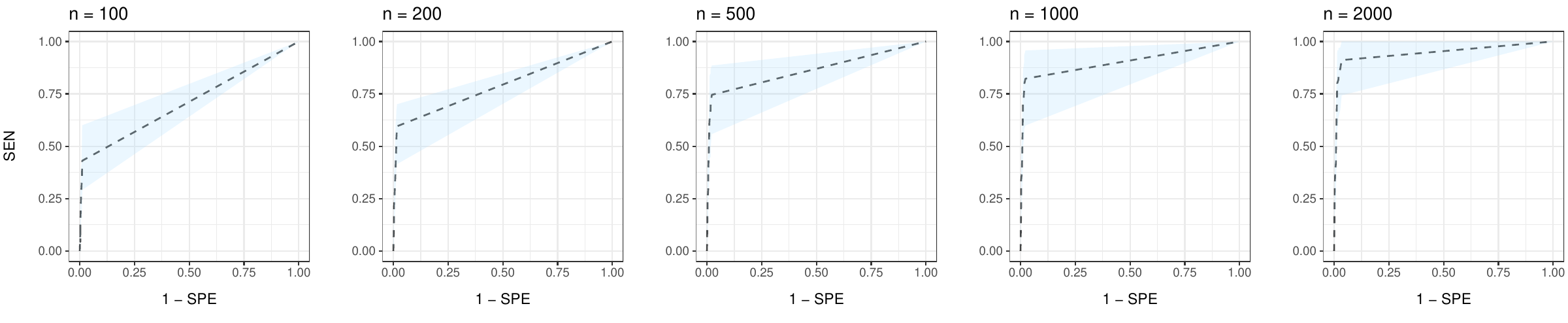}}
			\vspace{0.4cm}
			\\
			\multirow{1}{*}{\rotatebox[origin=c]{90}{\textit{Block \,\,\,\,\,\,  }}} & \raisebox{-0.7\height}{\includegraphics[scale=0.50]{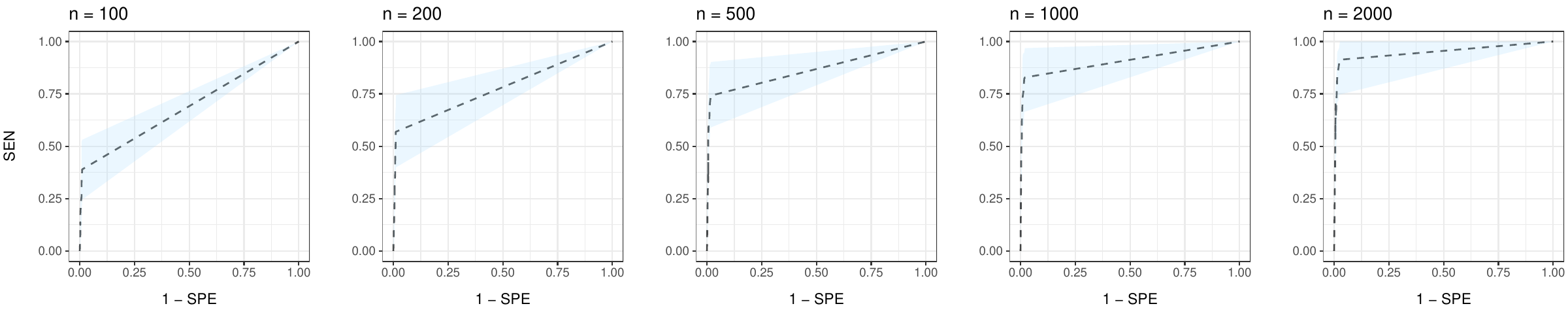}}
			%\end{figure}
		\end{tabular}
		\caption{\small Simulations. Receiver operating characteristic (ROC) curve obtained under varying thresholds for the posterior probabilities of edge inclusion for type of variables \textit{Mixed} $(b.4)$, sample size $n\in\{100,200,500,1000,2000\}$ $(c)$ and type of DAG structure \textit{Free, Regression, Block} $(a)$. Dotted lines represent the (average over the 40 simulated DAGs) ROC curve, while the blue area represents the 5th-95th percentile band.}
		\label{fig:sim:ROC:curves}
	\end{center}
\end{figure}
\end{landscape}

We now evaluate the performance of our method in recovering the true DAG structure.
To this end, we first estimate the posterior probabilities of edge inclusion in \eqref{eq:posterior:probability:edge:inclusion} for each pair of distinct nodes $u,v$.
Given a threshold for edge inclusion $k\in[0,1]$, we construct a graph estimate by including those edges $(u,v)$ such that $\widehat p_{u\rightarrow v} \ge k$.
We compare the resulting graph with the true DAG by computing the sensitivity (SEN) and specificity (SPE) indexes, defined as
\be
SEN = \frac{TP}{TP+FN}, \quad
SPE = \frac{TN}{TN+FP},
\ee
where $TP, TN, FP, FN$ are the numbers of true positives, true negatives, false positives and false negatives, based on the adjacency matrix of the estimated graph.
By varying the threshold $k$ within a grid in $[0,1]$ and computing SEN and SPE for each value of $k$, we obtain a
receiver operating characteristic (ROC) curve where each point corresponds to the values of $SEN$ and $(1-SPE)$ computed for a given threshold $k$.
Moreover, under each scenario, an \textit{average} (w.r.t.~the $N=40$ simulations) curve is constructed as follows.
For each threshold $k$, we compute $SEN$ and $(1-SPE)$ under each of the $40$ simulated DAGs and compute the average values of the two indexes. By repeating this procedure for each $k$ we obtain a collection of points which are joined by the average ROC curve. Results for Scenario \textit{Mixed} $(b.4)$, different sample sizes $(c)$ and types of DAG structures $(a)$ are summarized in Figure \ref{fig:sim:ROC:curves}.
We also proceed similarly to compute the 5th and 95th percentiles and obtain the blue band which is included in each plot of the same figure.
As expected, the performance of the method in recovering the true DAG structure improves as the sample size $n$ increases under all settings \textit{Free, Regression, Block}.

\begin{figure}
\begin{center}
	\begin{tabular}{lc}
		%\textit{Sparse Scenario} \quad\quad
		%\\
		\multirow{1}{*}{\rotatebox[origin=c]{90}{\textit{Free \quad \quad }}} & \raisebox{-0.7\height}{\includegraphics[scale=0.50]{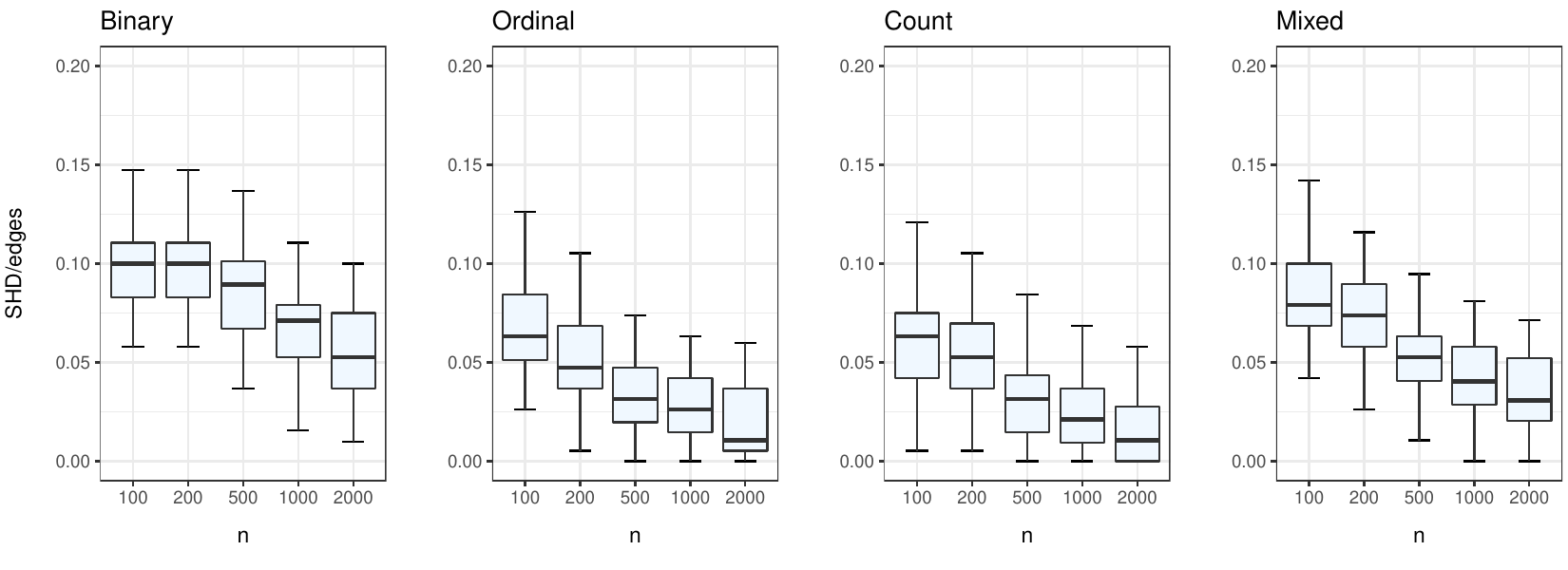}}
		\vspace{0.4cm}
		\\
		\multirow{1}{*}{\rotatebox[origin=c]{90}{\textit{Regression\quad }}} &
		\raisebox{-0.7\height}{\includegraphics[scale=0.50]{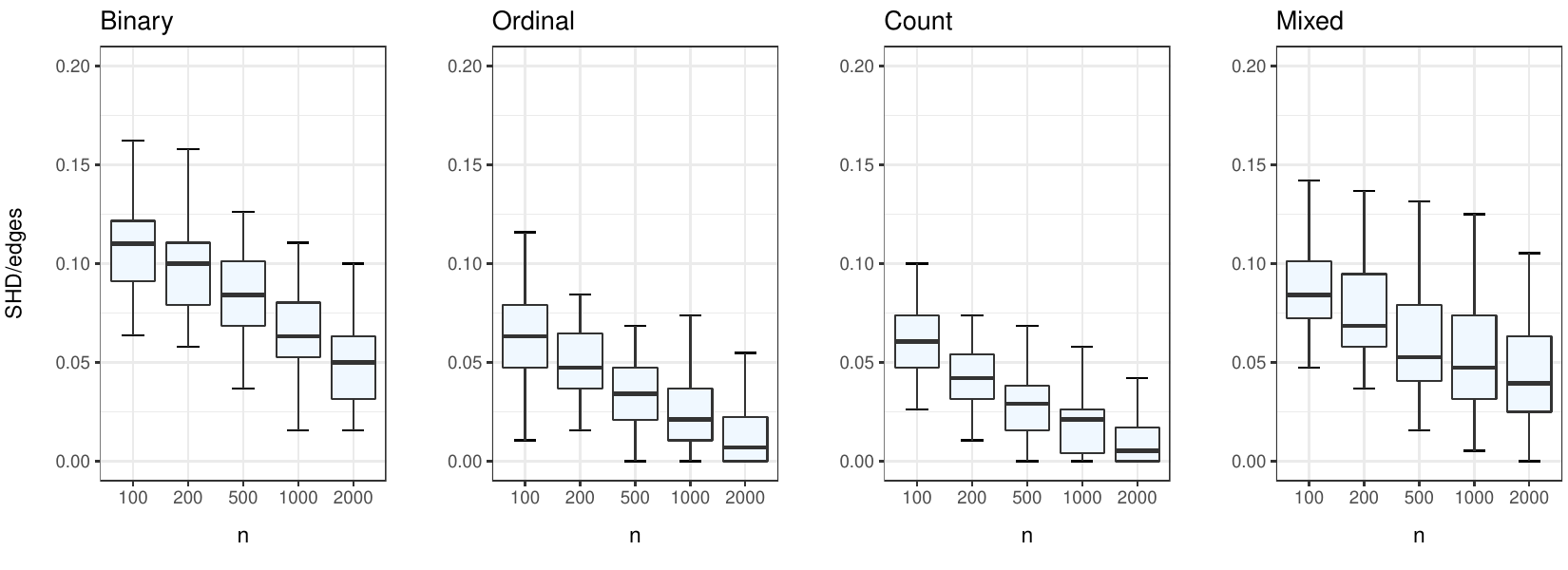}}
		\vspace{0.4cm}
		\\
		\multirow{1}{*}{\rotatebox[origin=c]{90}{\textit{Block \quad \,\, }}} & \raisebox{-0.7\height}{\includegraphics[scale=0.50]{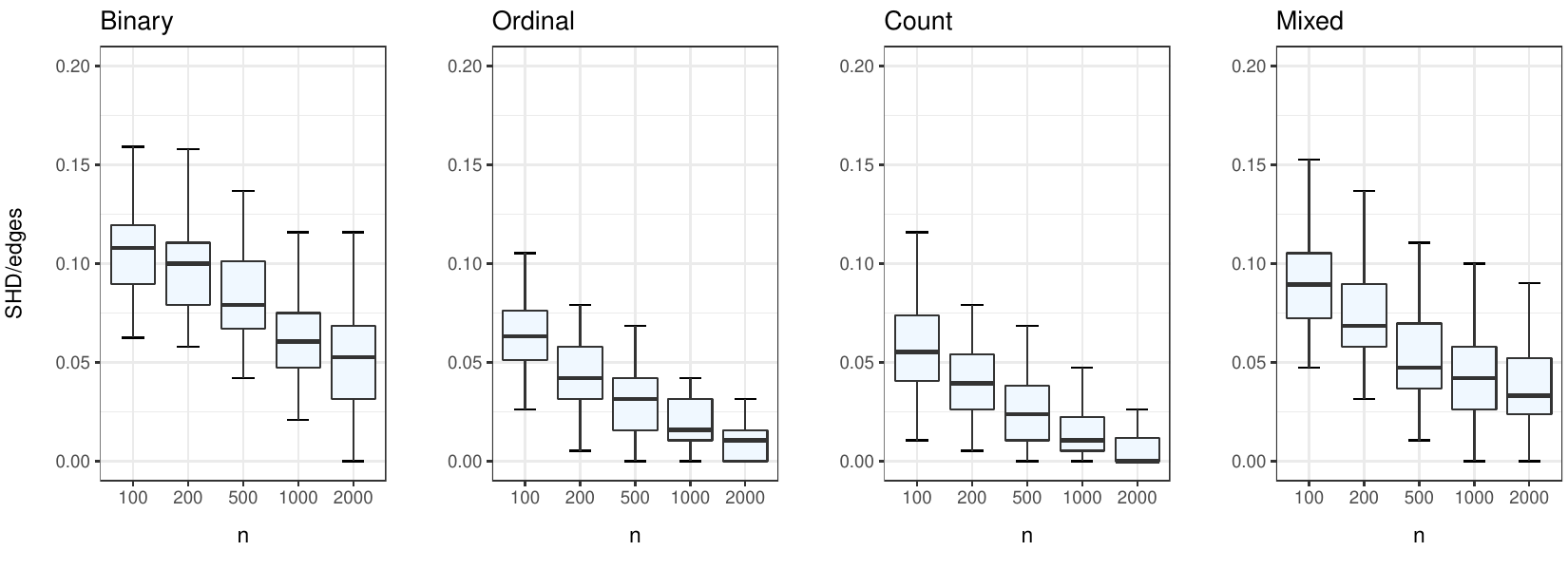}}
		%\end{figure}
	\end{tabular}
	\caption{\small Simulations. Distribution (across $N=40$ simulated datasets) of the relative Structural Hamming Distance (SHD/edges) between estimated and true DAG for type of variables \textit{Binary, Ordinal, Count, Mixed} $(b)$, sample size $n\in\{100,200,500,1000,2000\}$ $(c)$ and type of DAG structure \textit{Free, Regression, Block} $(a)$.}
	\label{fig:sim:SHD}
\end{center}
\end{figure}

As a single graph estimate summarizing the MCMC output we also consider the median probability (MPM) DAG model $\widehat{\D}$ which is obtained by fixing the threshold for edge inclusion as $k=0.5$.
We compare each DAG estimate $\widehat{\D}$ with the corresponding true DAG by measuring the Structural Hamming Distance (SHD). The latter corresponds to the number of edge insertions, deletions or flips needed to transform the estimated DAG into the true DAG; accordingly, lower values of SHD correspond to better performances.
Results for each simulation scenario are summarized in the box-plots of Figure \ref{fig:sim:SHD}, where each plot reports the distribution (across the $N=40$ simulated datasets) of the ratio between SHD and the maximum number of edges in the graphs (SHD/edges) for a combination of $(a)$ and $(c)$ and increasing sample sizes $n$.

The same behavior observed in Figure \ref{fig:sim:ROC:curves} for Scenario \textit{Mixed} is even more apparent,
with SHD decreasing at zero as $n$ increases under all settings.
In addition, DAG learning is more difficult in the \textit{Binary} scenario, where the collected categorical data provide a limited source of information to estimate dependence relations. By converse, structural learning is much easier under the \textit{Ordinal} and \textit{Count} scenarios. The performance in the case of mixed data is somewhat intermediate w.r.t.~the previous scenarios.

\begin{figure}
	\begin{center}
		\begin{tabular}{lc}
			%\textit{Sparse Scenario} \quad\quad
			%\\
			\multirow{1}{*}{\rotatebox[origin=c]{90}{\textit{Free \quad \quad }}} & \raisebox{-0.7\height}{\includegraphics[scale=0.50]{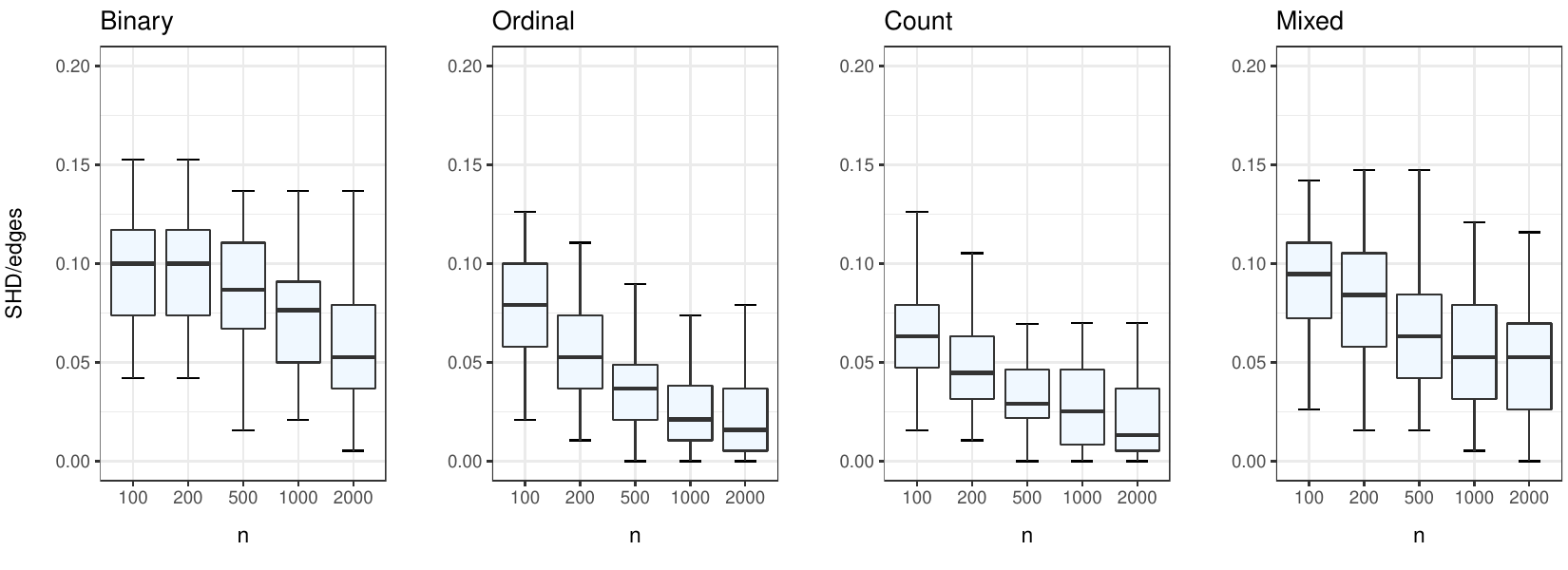}}
			\vspace{0.4cm}
			\\
			\multirow{1}{*}{\rotatebox[origin=c]{90}{\textit{Regression\quad }}} &
			\raisebox{-0.7\height}{\includegraphics[scale=0.50]{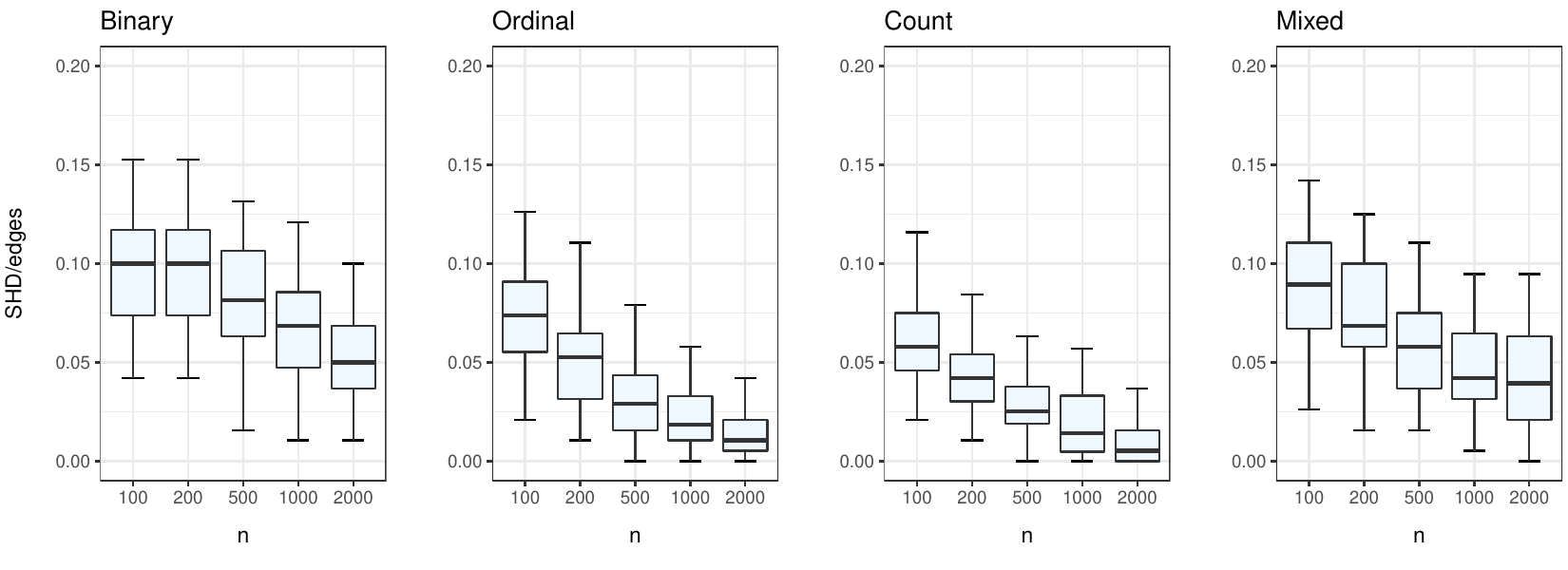}}
			\vspace{0.4cm}
			\\
			\multirow{1}{*}{\rotatebox[origin=c]{90}{\textit{Block \quad \,\, }}} & \raisebox{-0.7\height}{\includegraphics[scale=0.50]{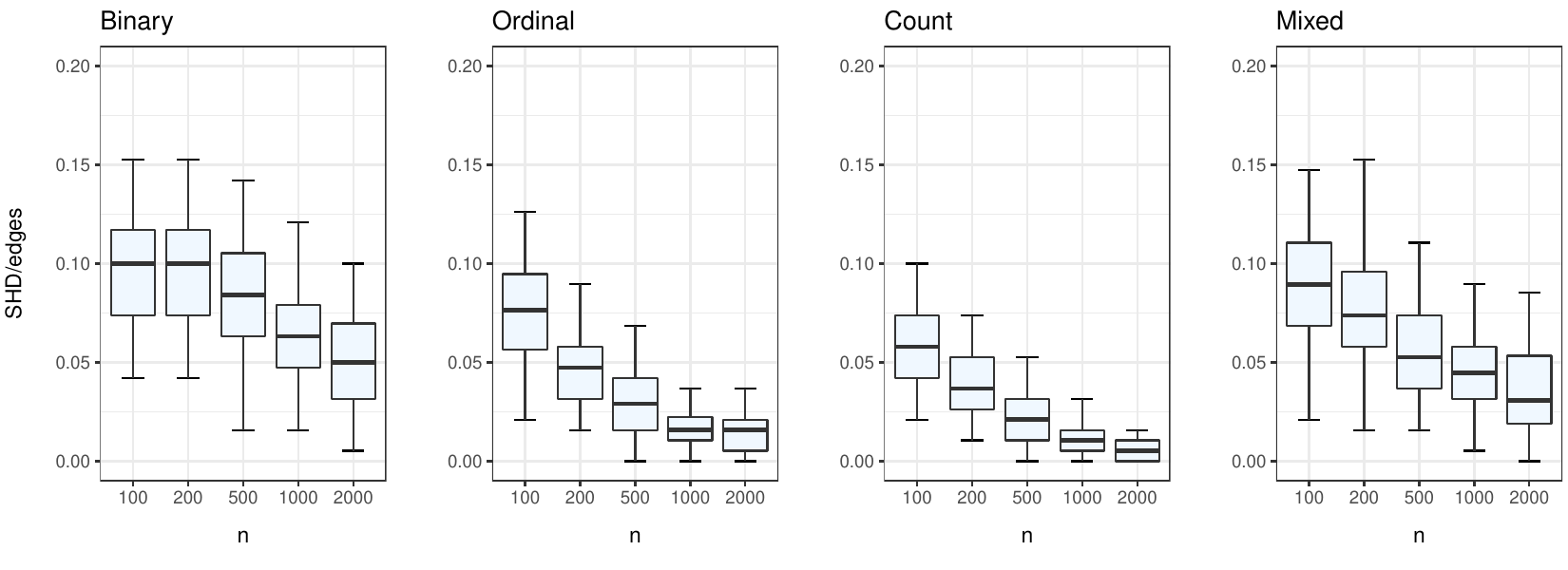}}
			%\end{figure}
		\end{tabular}
		\caption{\small Simulations. Distribution (across $N=40$ simulated datasets) of the relative Structural Hamming Distance (SHD/edges) between estimated and true DAG for type of variables \textit{Binary, Ordinal, Count, Mixed} $(b)$, sample size $n\in\{100,200,500,1000,2000\}$ $(c)$ and type of DAG structure \textit{Free, Regression, Block} $(a)$.}
		\label{fig:sim:SHD:unbalanced}
	\end{center}
\end{figure}

We finally summarize in Tables \ref{tab:sim:binary} - \ref{tab:sim:mixed}
the Sensitivity (SEN) and Specificity (SPE) indexes computed under the median probability DAG model estimate.
Each table refers to one scenario among \textit{Binary, Ordinal, Count, Mixed} and reports the average (w.r.t.~the $N=40$ simulations) percentage value of the two indexes under settings \textit{Free, Regression, Block} corresponding to the three different classes of DAG structures.
While SPE attains high levels even for small sample sizes, SEN significantly increases as $n$ grows. As a consequence, the improvement in the performance observed in Figure \ref{fig:sim:SHD} is mainly due to a reduction in the inclusion of ``false negative" edges as the number of available observations grows.
In addition, it appears that DAG identification is somewhat easier under Scenarios \textit{Regression} and \textit{Block}. Here, structural constraints limiting the DAG space can help identifying dependence relations between variables which otherwise may not be uniquely identifiable because of DAG Markov equivalence \citep{Ande:Madi:Perl:1997}.

\begin{table}
\caption{\small Simulations. Binary data. Specificity (SPE) and sensitivity (SEN) indexes (averaged over the 40 simulations) computed w.r.t.~the median probability DAG estimate, for sample size $n\in\{100,200,500,1000,2000\}$ and under each scenario \textit{Free, Regression, Block} corresponding to different classes of DAG structures.}
\centering
\small
\begin{tabular}{cccccccccc}
	\hline
	&& \multicolumn{2}{c}{Free} && \multicolumn{2}{c}{Regression} && \multicolumn{2}{c}{Block} \\
	$n$ && SPE & SEN && SPE & SEN && SPE & SEN \\
	\hline
	100 && 100.00 & 0.00 && 100.00 & 0.00 && 100.00 & 0.00 \\ 
	200 && 100.00 & 3.04 && 100.00 & 4.55 && 99.72 & 5.41 \\ 
	500 && 99.16 & 32.74 && 99.16 & 30.38 && 99.16 & 32.74 \\ 
	1000 && 98.87 & 54.07 && 98.88 & 53.85 && 99.02 & 56.52 \\ 
	2000 && 98.74 & 67.26 && 98.87 & 69.40 && 98.87 & 71.01 \\ 
	\hline
\end{tabular}
\label{tab:sim:binary}
\end{table}

\begin{table}
\caption{\small Simulations. Ordinal data. Specificity (SPE) and sensitivity (SEN) indexes (averaged over the 40 simulations) computed w.r.t.~the median probability DAG estimate, for sample size $n\in\{100,200,500,1000,2000\}$ and under each scenario \textit{Free, Regression, Block} corresponding to different classes of DAG structures.}
\centering
\small
\begin{tabular}{cccccccccc}
	\hline
	&& \multicolumn{2}{c}{Free} && \multicolumn{2}{c}{Regression} && \multicolumn{2}{c}{Block} \\
	$n$ && SPE & SEN && SPE & SEN && SPE & SEN \\
	\hline
	100 && 99.44 & 46.29 && 99.44 & 50.00 && 99.44 & 50.00 \\ 
	200 && 99.18 & 62.50 && 99.31 & 64.58 && 99.44 & 65.38 \\ 
	500 && 99.44 & 78.17 && 99.44 & 75.00 && 99.72 & 79.58 \\ 
	1000 && 99.43 & 85.10 && 99.44 & 87.50 && 99.72 & 89.06 \\ 
	2000 && 99.86 & 94.87 && 100.00 & 92.31 && 100.00 & 96.00 \\ 
	\hline
\end{tabular}
\label{tab:sim:ordinal}
\end{table}

\begin{table}
\caption{\small Simulations. Count data. Specificity (SPE) and sensitivity (SEN) indexes (averaged over the 40 simulations) computed w.r.t.~the median probability DAG estimate, for sample size $n\in\{100,200,500,1000,2000\}$ and under each scenario \textit{Free, Regression, Block} corresponding to different classes of DAG structures.}
\centering
\small
\begin{tabular}{cccccccccc}
	\hline
	&& \multicolumn{2}{c}{Free} && \multicolumn{2}{c}{Regression} && \multicolumn{2}{c}{Block} \\
	$n$ && SPE & SEN && SPE & SEN && SPE & SEN \\
	\hline
	100 && 99.43 & 50.00 && 99.43 & 50.00 && 99.43 & 54.07 \\ 
	200 && 99.16 & 66.67 && 99.44 & 69.23 && 99.44 & 69.80 \\ 
	500 && 99.44 & 78.71 && 99.44 & 80.00 && 99.72 & 82.09 \\ 
	1000 && 99.45 & 88.12 && 99.72 & 88.87 && 99.86 & 91.11 \\ 
	2000 && 100.00 & 95.74 && 100.00 & 98.33 && 100.00 & 100.00 \\ 
	\hline
\end{tabular}
\label{tab:sim:count}
\end{table}

\begin{table}
\caption{\small Simulations. Mixed data. Specificity (SPE) and sensitivity (SEN) indexes (averaged over the 40 simulations) computed w.r.t.~the median probability DAG estimate, for sample size $n\in\{100,200,500,1000,2000\}$ and under each scenario \textit{Free, Regression, Block} corresponding to different classes of DAG structures.}
\centering
\small
\begin{tabular}{cccccccccc}	
	\hline
	&& \multicolumn{2}{c}{Free} && \multicolumn{2}{c}{Regression} && \multicolumn{2}{c}{Block} \\
	$n$ && SPE & SEN && SPE & SEN && SPE & SEN \\
	\hline
	100 && 99.45 & 23.74 && 99.44 & 25.00 && 99.43 & 27.92 \\ 
	200 && 98.90 & 37.72 && 99.02 & 36.67 && 99.16 & 45.99 \\ 
	500 && 98.88 & 61.39 && 98.87 & 62.50 && 99.16 & 65.94 \\ 
	1000 && 98.89 & 71.01 && 98.86 & 73.68 && 99.01 & 74.46 \\ 
	2000 && 98.30 & 78.71 && 98.58 & 80.00 && 98.57 & 84.11 \\ 
	\hline
\end{tabular}
\label{tab:sim:mixed}
\end{table}

\subsection{Simulation experiments with unbalanced correlation structure}

In the simulation scenarios considered before we randomly drew the non-zero elements of matrix $\bL$, corresponding to regression coefficients in the latent linear SEM, uniformly in the symmetric interval $[-1,-0.1]\cup[0.1,1]$. This implies an expected ``balanced" correlation structure between variables, meaning that both
positive and negative associations in the generating
model will be present, and with same expected proportion.
However, in many psychological applications variables are mostly positively correlated each other; see also our results in Section 6. An important example is represented by cognitive test scores, whose pattern of positive correlations was already advocated by \citet{Spearman:1904} in his positive manifold theory. To assess the ability of our method to capture such ``unbalanced" correlation structure, we now consider a random choice of the non-zero elements of $\bL$ in the interval $[0.1,1]$ and implement
the same scenarios introduced in Section 5.1. Accordingly, all variables are now positively correlated at the latent level.
Results, in terms of relative SHD between true and estimated DAG are reported in Figure \ref{fig:sim:SHD:unbalanced}, which summarizes the distribution of SHD across simulated datasets under all scenarios defined in Section 5.1. The performance of our method is in line with
what observed in Figure \ref{fig:sim:SHD}, obtained under a balanced setting, suggesting that our model specification is
insensitive to the specific proportion of positive/negative correlations.

\black

\section{Real data analyses}
\label{sec:application}

\subsection{Well-being data}
\label{sec:application:well:being}

%In this section we apply the proposed methodology to two different real-data problems. In particular, we first analyze survey data collected from a social study on well being and then consider a medical dataset relative to patients affected by cardiovascular disease.

%\subsection{Well-being data}

%We refer to the web link for full details on the data.

The \textit{Your Work-Life Balance} is a project promoted by the United Nations (UN) and implemented through a public survey available at \url{http://www.authentic-happiness.com/your-life-satisfaction-score}.
Scope of the survey is to evaluate how people
thrive in both professional and personal lives based on several indicators that are related with life satisfaction.
Variables considered in the study are classified into five dimensions:
\begin{itemize}
\item Healthy body, features reflecting fitness and healthy habits;
\item Healthy mind, indicating how well subjects embrace positive emotions;
\item Expertise, measuring the ability to grow expertise and achieve something unique;
\item Connection, assessing the strength of social relationships;
\item Meaning, evaluating compassion, generosity and happiness.
\end{itemize}
%We also refer to the web link for more details.
%
The survey supports the UN Sustainable Development Goals
(\url{https://sdgs.un.org})
and aims at providing insights on the determinants of human well-being.
Accordingly, some questions of interest are the following:
\begin{itemize}
	\item ``What are the strongest correlations between the various dimensions?"
	\item ``What are the best predictors of a balanced life?"
\end{itemize}
%This study supports the following UN Sustainable Development Goals:
%8.4 Improve global resource efficiency in consumption and production and endeavour to decouple economic growth from environmental degradation
%12.8 Ensure that people everywhere have the relevant information and awareness for sustainable development and lifestyles in harmony with nature
%12.8.1 Extent to which (i) global citizenship education and (ii) education for sustainable development (including climate change education) are mainstreamed in (a) national education policies; (b) curricula; (c) teacher education; and (d) student assessment

The complete dataset is publicly available at \url{https://www.kaggle.com/datasets/ydalat/lifestyle-and-wellbeing-data}.
It includes observations collected across years
$2015-2020$ of $20$ ordinal variables (with levels ranging in $1-5$ or $1-10$) each measuring
closeness of a subject w.r.t. to one perceived dimension, besides gender (binary) and age (ordinal with four classes).
We include in our analysis the $n = 459$ observations available for year $2020$.
We consider variable \textit{stress} as the response and accordingly allow edges \textit{from} each of the remaining variables \textit{to} the response only. In addition, \textit{age} and \textit{gender} are considered as objective features and accordingly cannot have incoming edges. We do not impose further constraints among the remaining variables in terms of edge directions that are known in advance.

Following the scope of the original project, we are interested in understanding how the various dimensions relate each other and what are the direct/indirect determinants of the perceived level of stress.
To this end, we implement our MCMC algorithm for a number of iterations $S=20000$, after an initial burnin period of $5000$ runs that are discarded from posterior analysis. Diagnostic tools based on multiple chains and graphical inspections of the behavior across iterations of sampled parameters were also adopted to assess the convergence of the MCMC; see Section \ref{appendix:diagnostics} for details.

\begin{figure}
\begin{center}
	\begin{tabular}{c}
		\includegraphics[scale=0.44]{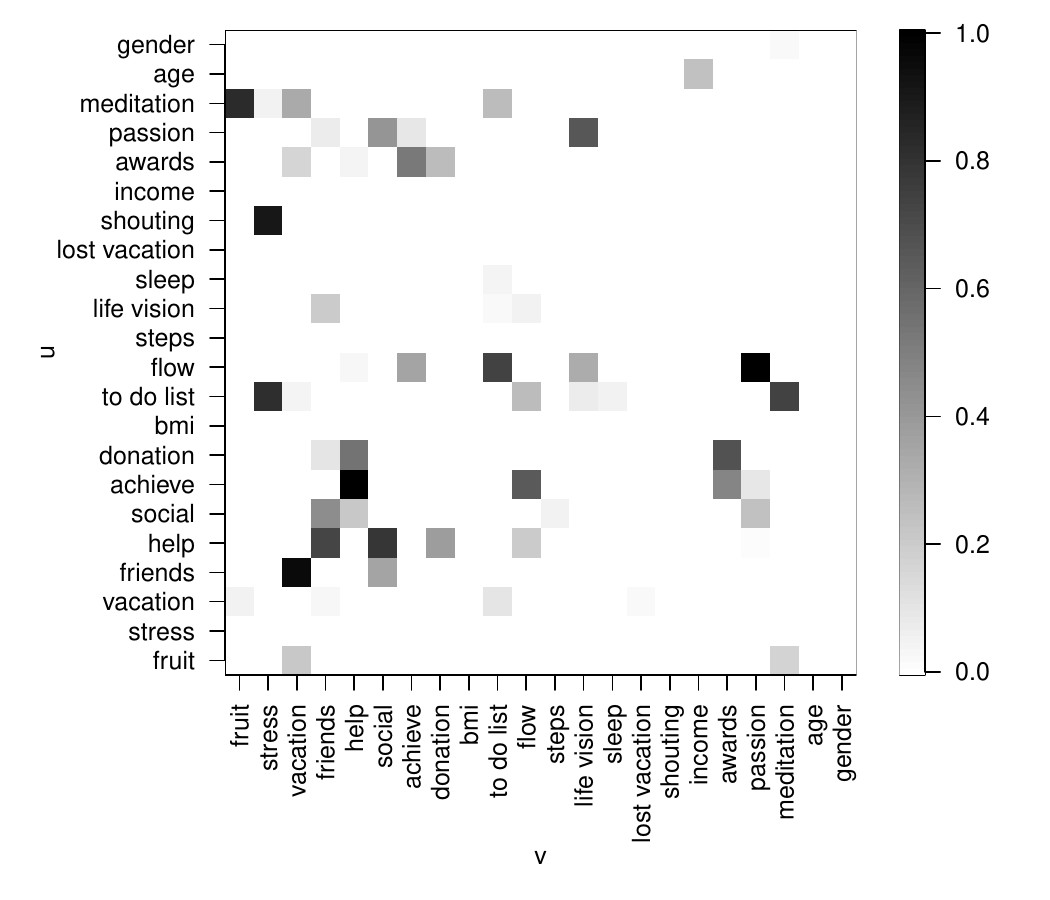}
		\includegraphics[scale=0.44]{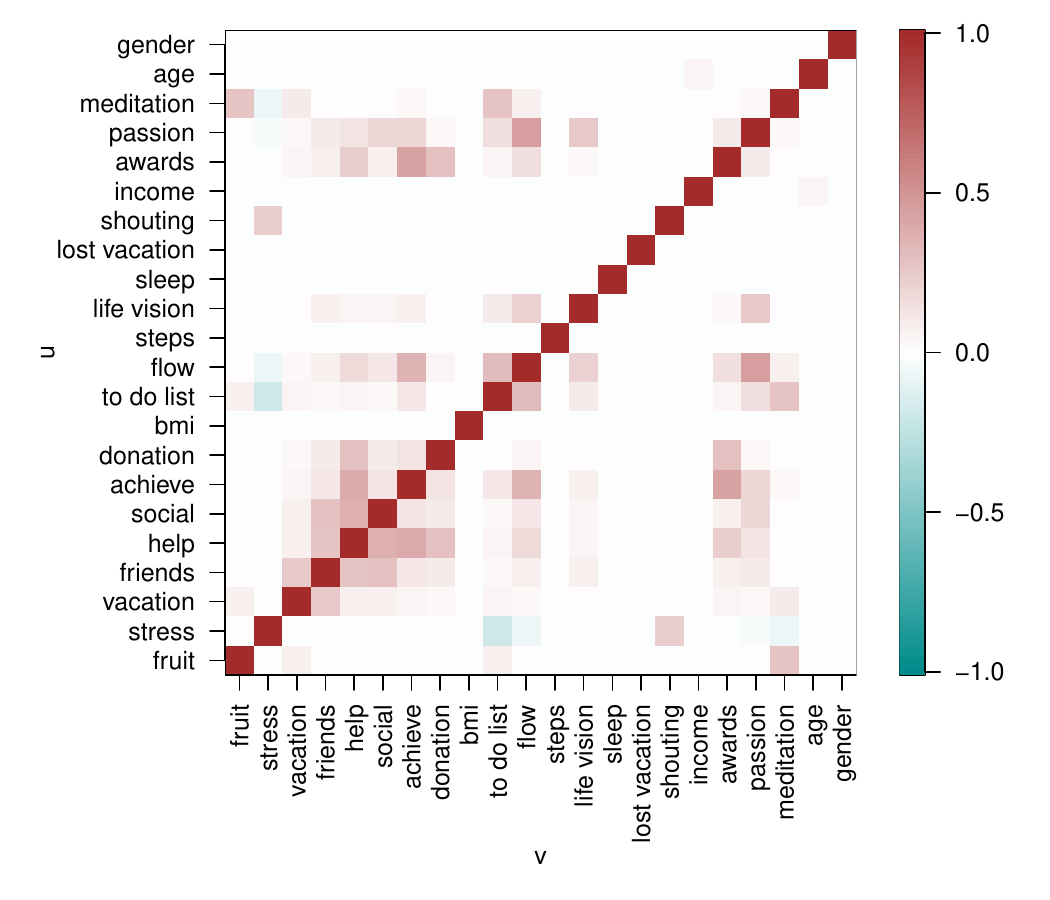}
		%\end{figure}
	\end{tabular}
	\caption{\small Well-being data. Upper panel: Heat map with estimated posterior probabilities of edge inclusion $\widehat p (u\rightarrow v \g \bX)$ for each edge $(u,v)$. Lower panel: Estimated correlation matrix.}
	\label{fig:well:being:maps}
\end{center}
\end{figure}

\begin{figure}
	\begin{center}
		\begin{tabular}{lc}
			\includegraphics[scale=0.64]{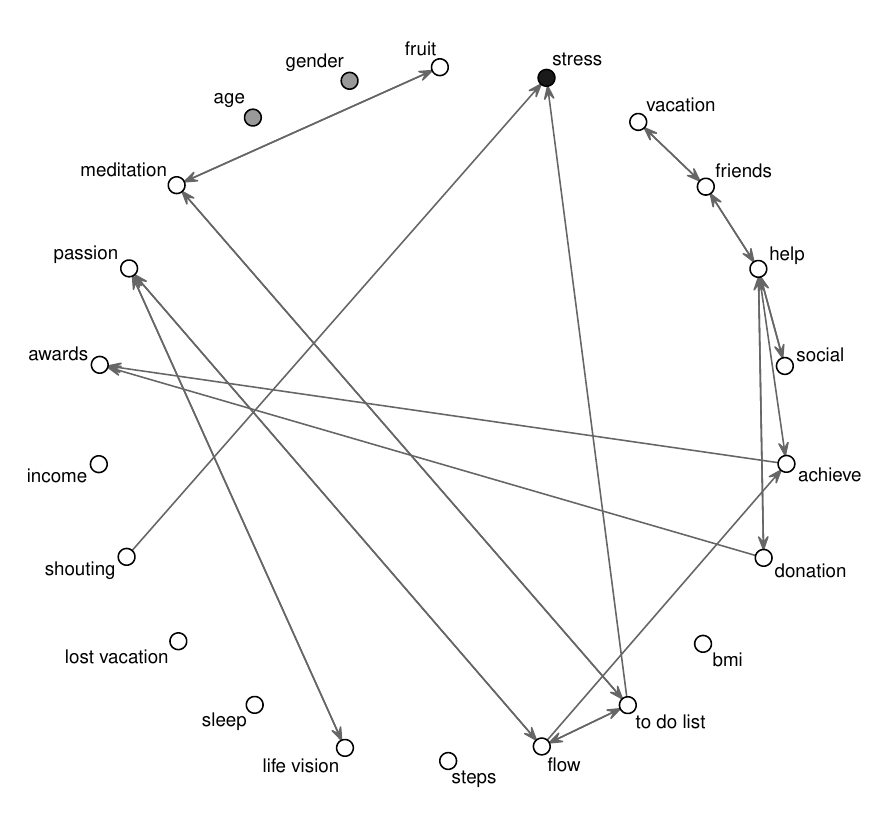}
			%\end{figure}
		\end{tabular}
		\caption{\small Well-being data. Estimated CPDAG.}
		\label{fig:well:being:dag}
	\end{center}
\end{figure}

We use the MCMC output to provide an estimate of the posterior probabilities of edge inclusion as well as a BMA estimate of the correlation matrix between (latent) variables. Results are reported in the two heat maps of Figure \ref{fig:well:being:maps}.
The upper map, which collects the estimated posterior probabilities of edge inclusion clearly suggests a sparse structure in the underlying network, with only a few edges whose posterior probability exceeds $0.5$. This also emerges from the graph estimate reported in Figure \ref{fig:well:being:dag}, which corresponds to the CPDAG representing the equivalence class of the MPM DAG estimate.
%In addition, the right-hand side of Figure \ref{fig:well:being:maps} reports the estimated correlation matrix.
%
\black

%IN A TYPICAL DAY, HOW MANY HOURS DO YOU EXPERIENCE "FLOW"?
%*
%► 'Flow' is defined as the mental state, in which you are fully immersed in performing an activity. You then experience a feeling of energized focus, full involvement, and enjoyment in the process of this activity. ►Watch the youtube video from Mihaly Csikszentmihalyi "Flow, the secret to happiness"

The estimated graph reveals that two variables directly affect the perceived level of stress, namely \textit{shouting} (\textit{``How often do you shout or sulk at somebody?"}) and \textit{to do list} (\textit{``how well
do you complete your weekly to-do list?"}).
Moreover, variable \textit{stress} is conditionally independent from \textit{flow} (\textit{``How many hours you experience flow, i.e. you fell fully immersed in performing an activity?"}) given \textit{to do list}. Also, \textit{flow} and \textit{to do list} are positively correlated, which suggests that people performing better in their activities also follow through with many more of their weekly goals. This in turn has a direct impact on the perceived level of stress.
It also appears that \textit{shouting} is positively correlated with the level of stress, while there is negative correlation with \textit{to do list}; accordingly
better results in completing to-do-lists, imply a reduction in the stress level perceived by individuals.

\subsection{Student mental health data}
\label{sec:application:medical}

We consider a dataset from a cross-sectional study conducted on $886$ medical students in Switzerland and presented by \citet{Carrad:et:al:2022:mental}.
Target of the study is to provide insights on students' well-being, in order to implement policies aimed at improving their academic-life satisfaction and conditions.
A number of dimensions related to empathy are measured through self-reported questionnaires based on the Questionnaire of Cognitive and Affective Empathy (QCAE) and
the Jefferson Scale of Physician Empathy (JSPE); %emotion recognition tests
related variables are the QCAE affective empathy score (\texttt{qcae aff}),
the QCAE cognitive empathy score (\texttt{qcae cog}) and the JSPE total empathy score (\texttt{jspe}).
Burnout is a state of emotional, physical, and mental exhaustion which is caused by excessive exposure to stress. The burnout dimension is measured through the Maslach Burnout Inventory-Student Survey (MBI-SS); the latter is based on $15$ items and provides three scores evaluating the following dimensions: emotional exhaustion (\texttt{mbi ex}), cynicism (\texttt{mbi cy}), and academic efficacy (\texttt{mbi ea}).
In addition, students' anxiety and depression is measured
through the Center for Epidemiologic Studies Depression (CESD) score
and the State-Trait Anxiety Inventory (STAI) score, both based on a questionnaire with self-report items on Likert scales (\texttt{cesd} and \texttt{stai} respectively).
%Higher scores are positively correlated with higher levels of anxiety.
%The State-Trait Anxiety Inventory (STAI) is a psychological inventory consisting of 40 self-report items on a 4-point Likert scale. The STAI measures two types of anxiety – state anxiety and trait anxiety. Higher scores are positively correlated with higher levels of anxiety.
Finally, the dataset contains information on demographic factors such as \texttt{age} and \texttt{gender}, besides
variables measuring job satisfaction (\texttt{job}),
%psychological distress, education grades
partnership status (\texttt{part})
and self-reported health status (\texttt{health}), represented by an ordinal variable with $5$ categories corresponding to increasing levels of perceived health satisfaction.
We refer to \citet{Carrad:et:al:2022:mental} for a detailed description of the complete dataset, which is publicly available at \url{https://zenodo.org/record/5702895}.
We emphasize that the structure of the analyzed dataset is quite heterogeneous, as it collects binary, ordinal (with different ranges of levels) as well as continuous measurements simultaneously.

%health	SATISFACTION WITH HEALTH : How satisfied are you with your health?	1=Verydissatisfied; 2=Dissatisfied; 3=Neithersatisfiednordissatisfied; 4=Satisfied; 5=Verysatisfied
%
One specific aim of the original study is to identify how variables included in the survey, in particular depressive symptoms, anxiety, and burnout, are related to empathy and mental health.
In our analysis, we consider \texttt{age}, \texttt{sex}, \texttt{part} and \texttt{job} as exogenous variables, while we regard \texttt{health} as a response of interest.

\begin{figure}
	\begin{center}
		\begin{tabular}{c}
			\includegraphics[scale=0.44]{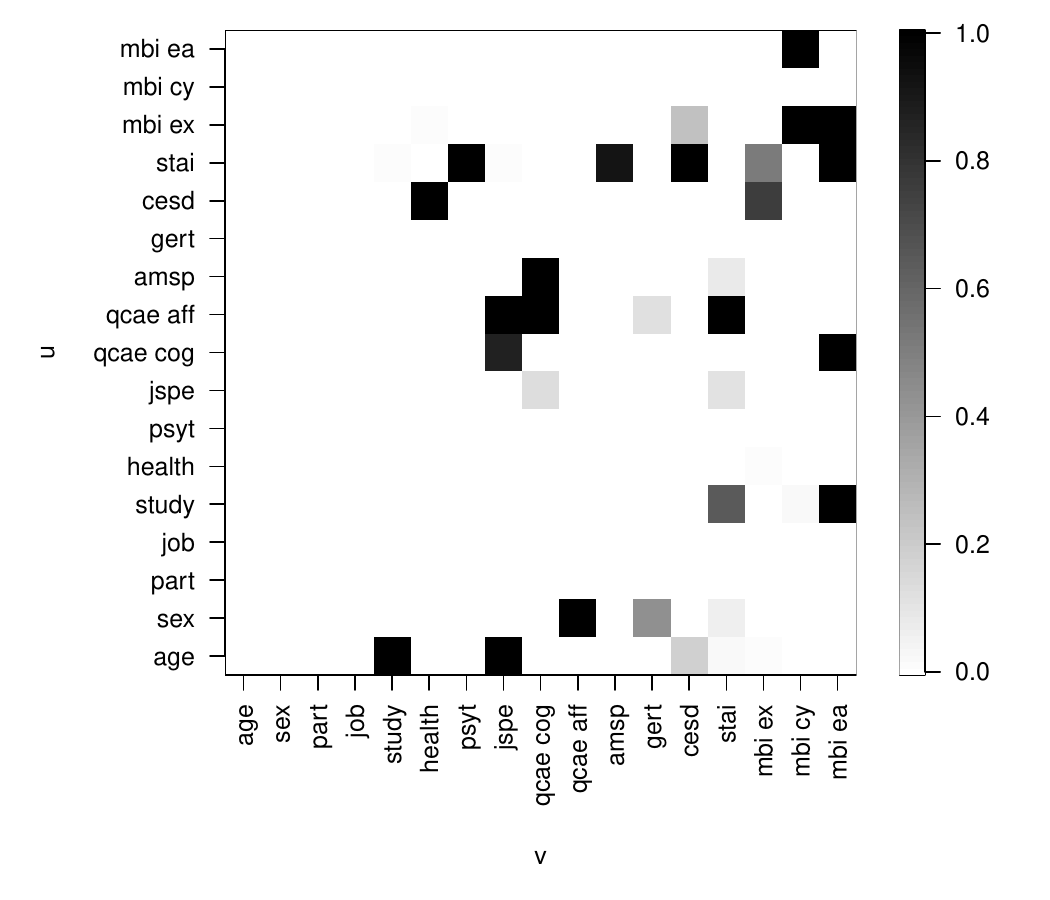}
			\includegraphics[scale=0.44]{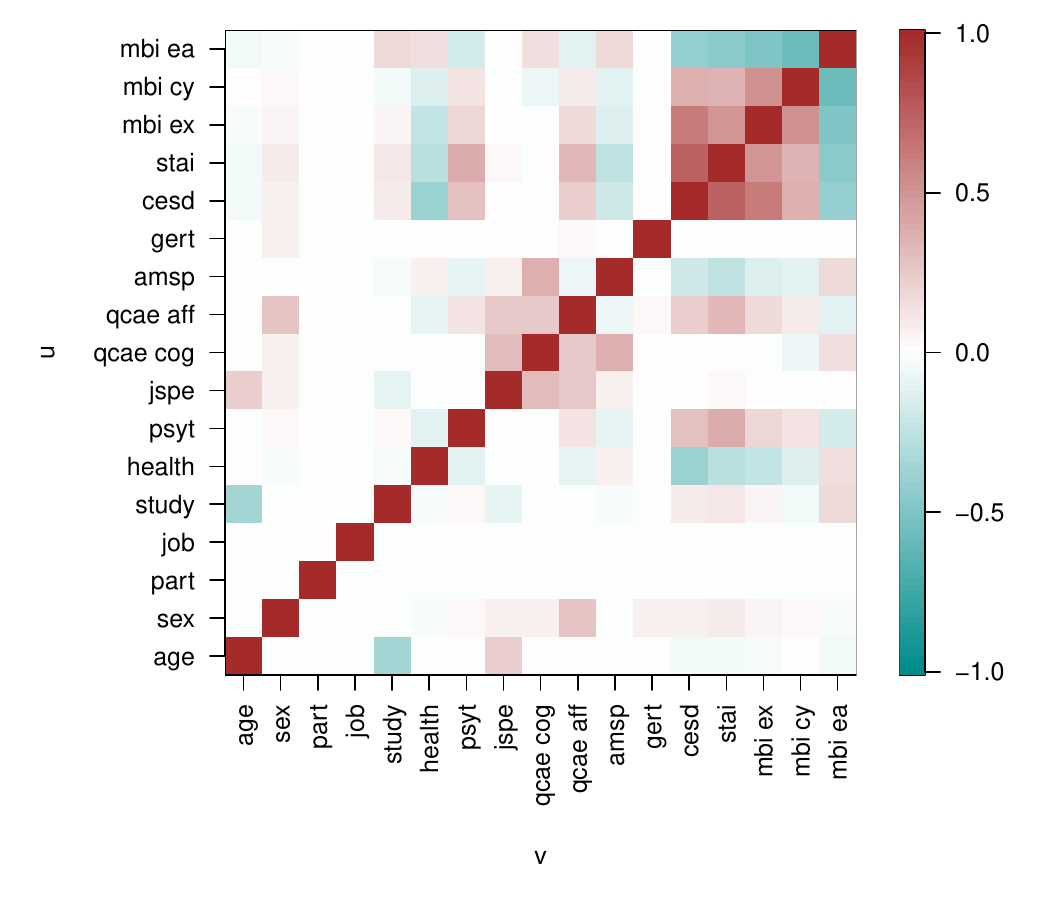}
			%\end{figure}
		\end{tabular}
		\caption{\small Student mental health data. Upper panel: Heat map with estimated posterior probabilities of edge inclusion $\widehat p (u\rightarrow v \g \bX)$ for each edge $(u,v)$. Lower panel: Estimated correlation matrix.}
		\label{fig:mentall:maps}
	\end{center}
\end{figure}

\begin{figure}
	\begin{center}
		\begin{tabular}{lc}
			\includegraphics[scale=0.75]{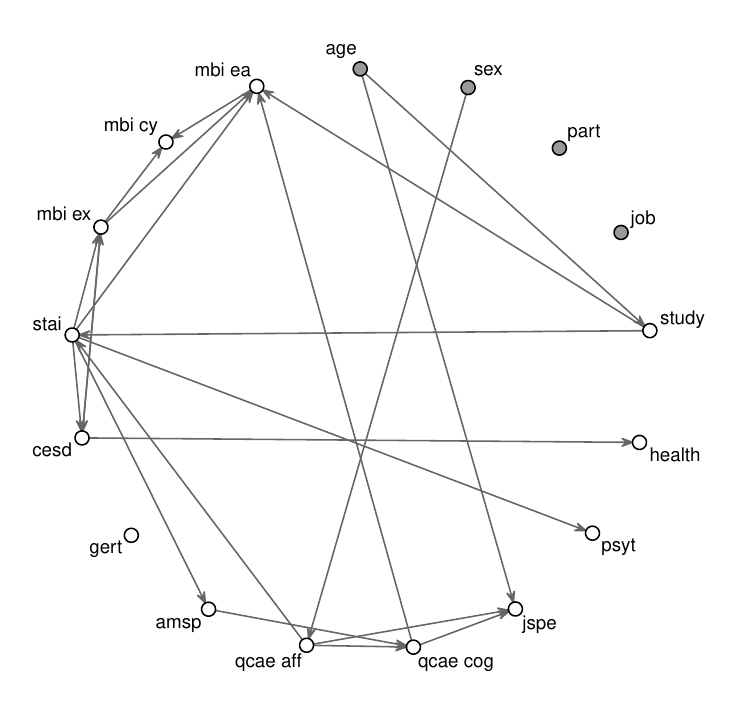}
			%\end{figure}
		\end{tabular}
		\caption{\small Student mental health data. Estimated CPDAG.}
		\label{fig:mental:dag}
	\end{center}
\end{figure}

Our MCMC algorithm is implemented for $S=20000$ iterations, after a burnin period of $5000$ runs that we adopt to assess the convergence of the chain.
We use the MCMC output to provide an estimate of the (marginal) posterior probability of inclusion (PPI)
for each possible directed edge in the DAG space; see Equation \eqref{eq:posterior:probability:edge:inclusion}.
The resulting heat-map with the collection of estimated PPIs (Figure \ref{fig:mentall:maps}, upper plot) suggests %strong evidence relative to
the existence of a few strong dependence relations among variables that correspond to directed links and paths in the graphs visited by the MCMC chain; this also appears from the MPM CPDAG estimate reported in Figure \ref{fig:mental:dag} which contains a moderate number of edges.
Furthermore, to investigate how variables
%that are connected in the graph
correlate each other, we provide a BMA estimate of the correlation matrix between (latent) variables (Figure \ref{fig:mentall:maps}, lower plot).

Notably, most variables belonging to the burnout dimension as well as to the
anxiety or depression sphere are positively correlated
with the exception of \texttt{mbi ea}, namely the score measuring academic efficacy, which as expected is negatively associated with the other variables, and in particular with the perceived level of anxiety (\texttt{stai}). \texttt{mbi ea} is also positively influenced by
variable \texttt{study} (hours of study \textit{per} week), which in turn has a positive, although less marked, effect on student's anxiety.
% study -> stai with positive correlation
% study -> mbi ea with positive correlation (più studio, meglio performo accademicamente, ovviamente)
Importantly however, students' depression, here quantified by the \texttt{cesd} score, has
a negative effect on \texttt{health},
as it appears from the graph estimate of Figure \ref{fig:mental:dag} and the correlation matrix in Figure \ref{fig:mentall:maps}.
Also, \texttt{age} has a positive impact on variable \texttt{jspe} which summarizes the empathy dimension, implying that medical students develop throughout their academic life a growing ability
in understanding and sharing feelings that are experienced by others.
Finally, an interesting set of structural dependencies is represented by the directed path \texttt{study} $\rightarrow$ \texttt{stai} $\rightarrow$ \texttt{cesd} $\rightarrow$ \texttt{health}.
The latter structure suggests that students more involved in studying activities may incur in higher levels of anxiety and depression, which consequently affects personal health conditions. 

\black

\section{Discussion}
\label{sec:discussion}

We proposed a Bayesian semi-parametric methodology for structure learning of directed networks which applies to mixed data, i.e.~data that include categorical, discrete and continuous measurements.
Our model formulation assumes that the available observations are generated by latent variables whose joint distribution is multivariate Gaussian and satisfies the independence constraints imposed by a Directed Acyclic Graph (DAG).
Following a copula-based approach, we model separately the dependence parameter and the marginal distributions of the observed variables which are estimated non-parametrically. The former corresponds to the covariance matrix, Markov w.r.t.~an unknown DAG, for which a DAG-Wishart prior is assumed. Importantly, we fully account for model uncertainty by assigning a prior distribution to DAG structures. In addition, constraints on the network structure that are known beforehand can be easily incorporated in our model. The resulting framework allows for posterior inference on DAGs and DAG-parameters which is carried out by implementing an MCMC algorithm.
We finally investigate the performance of our methodology through simulation studies. Results show that our method, when adopted to provide a single network estimate is highly competitive with the frequentist benchmark Copula PC.
In addition, being Bayesian, uncertainty around network structures as well as dependence parameters is fully provided by our method.

The theory of latent traits assumes that available observations collected on subjects are associated to a (possibly small) number of latent individual characteristics \citep{Hambleton:Cook:latent:traits:1977}. A latent trait model then establishes a mathematical relationship between these unobservable features and the observed data, which represent manifestations of the traits.
In this context, \citet{Moustaki:Knott:2000} generalized the classical latent trait model, originally introduced for categorical manifest variables, to mixed data and specifically variables whose distribution belongs to the exponential family. For each manifest variable they specify a suitable generalized linear model, where covariates correspond to a set of common latent traits following independent normal distributions.
Differently, our graphical modelling framework employs latent variables to project observable mixed-type features into a latent space (of same dimension) which, once equipped with a network model, incorporates a dependence structure between latent variables.
We conjecture that our method could be extended to a latent trait framework for mixed data as the one considered by \citet{Moustaki:Knott:2000}. A related model formulation would establish a link between each of the $q$ manifest variables and $K<q$ latent factors whose joint Gaussian distribution satisfies the conditional independencies embedded in a DAG. Finally, the method could identify a set of dependence relations between latent traits represented through a directed network.
%Those types of latent variables models employ a smaller number (say K)
%of latent variables/factors than the number of measurements to model the dependency
%of multivariate responses via a q × K loading matrix.
\black

Our model formulation assumes a common DAG structure with allied dependence parameter for all the available observations. \black
In some settings however, a \emph{clustering} structure may be present in the sample, with subjects divided into groups that are defined beforehand or unknown and therefore to learn from the data.
In the former case, multiple datasets could be analyzed jointly by using a multiple graphical model approach \citep{Peterson:et:al:2015:JASA}. The latter adopts a Markov random field prior that encourages common edges among group-specific graphs, and a spike-and-slab prior controlling network relatedness parameters.
In the second case, one could instead set up a mixture model,
where each mixture component corresponds to a possibly different network with allied parameters; as the output, a clustering structure of the subjects would be also available; see for instance \citet{Ickstadt:et:al:2010} and \citet{Lee:et:al:2022} for mixtures of graphical models in the Gaussian and ordinal framework respectively.
Following a Bayesian non-parametric approach we could consider an infinite mixture model where each latent group is characterized by a component-specific parameter.
A Dirichlet Process (prior) on the space of DAGs and DAG-parameters could be then assumed; see in particular \citet{Rodriguez:et:al:2011:EJS} and \citet{Castelletti:Consonni:2023:bnp} for respectively undirected and directed Gaussian graphical models.
Extensions of the proposed copula model to multiple DAGs and mixtures of DAGs are possible and currently under investigation.
%In both cases a different network structure would be available for each group.
%For example, with regard to our first applied problem, this could help emphasizing differences as well as similarities among subjects
%receiving different medical treatments or belonging to
%different sub-populations such as countries.

\newpage

\section*{Appendix}
\label{sec:appedix}

\subsection*{Comparison with Copula PC}

In this section we compare our methodology with the benchmark Copula PC method of \citet{Cui:et:al:2016:PC:copula}; see also \citet{Cui:et:al:2018}.
Copula PC is a two-step approach which can be applied to mixed data comprising categorical (binary and ordinal), discrete and continuous variables.
It first
estimates a correlation matrix in the space of latent variables (each associated with one of the observed variables) which is then used to test conditional independencies as in the standard PC algorithm.
For the first step, the same Gibbs sampling scheme introduced by \citet{Hoff:2007} and based on data augmentation with latent observations is adopted.
Moreover, conditional independence tests are implemented at significance level $\alpha$ which we vary in $\{0.01,0.05,0.10\}$; lower values of $\alpha$ imply a higher expected level of sparsity in the estimated graph. We refer to the three benchmarks as Copula PC 0.01, 0.05 and 0.10 respectively.
Output of Copula PC is a Completed Partially Directed Acyclic Graph (CPDAG) representing the estimated equivalence class.
With regard to our method, we also consider as a single graph estimate summarizing our MCMC output the CPDAG representing the equivalence class of the estimated median probability DAG model.
Each model estimate is finally compared with the true CPDAG by means of the SHD between the two graphs.

Results for Scenario \textit{Free}, type of variables \textit{Binary, Ordinal, Count, Mixed} and each sample size $n\in\{100,200,500,1000,2000\}$ are summarized in Figure \ref{fig:sim:cfr} which reports the distribution across $N=40$ simulations of the SHD.
It first appears that all methods improve their performances as the sample size $n$ increases. In addition, structure learning is more difficult in the \textit{Binary} case, while easier in general in the case of \textit{Ordinal} and \textit{Count} data. Moreover, Copula PC 0.01 (light grey) performs better than Copula PC 0.05 and 0.10 (middle and dark gray respectively). Our method clearly outperforms the three benchmarks in the \textit{Binary} scenario, a behavior which is more evident for large sample sizes. In addition, it performs better than Copula PC 0.05 and 0.10 most of the time under the remaining settings and remains highly competitive with Copula PC 0.01, with an overall better performance in terms of average SHD under almost all sample sizes for Scenarios \textit{Ordinal} and \textit{Count}.

\begin{landscape}
	\begin{center}
		\begin{figure}
			\begin{center}
				\begin{tabular}{cc}
					%\textit{Sparse Scenario} \quad\quad
					%\\
					\includegraphics[scale=0.56]{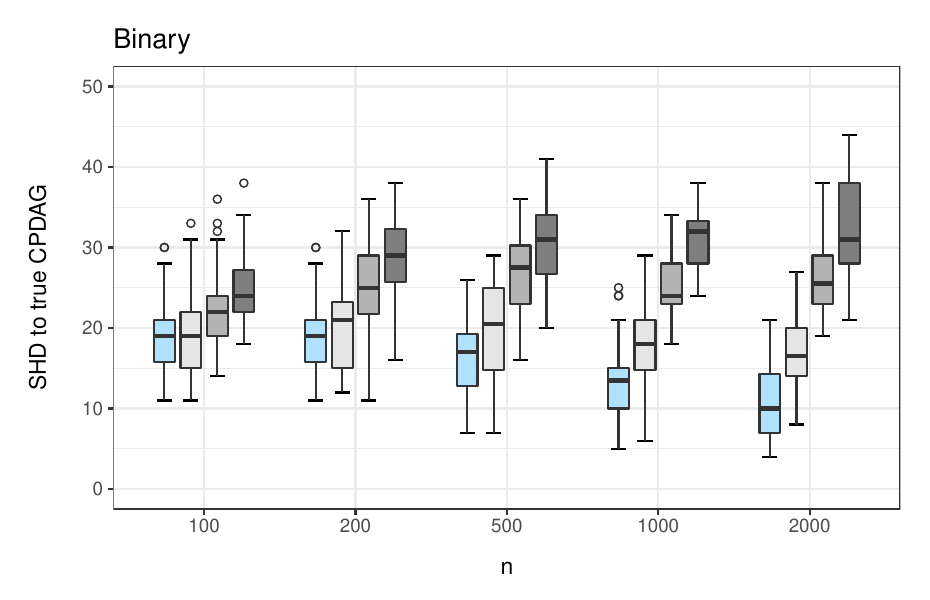}
					&
					\includegraphics[scale=0.56]{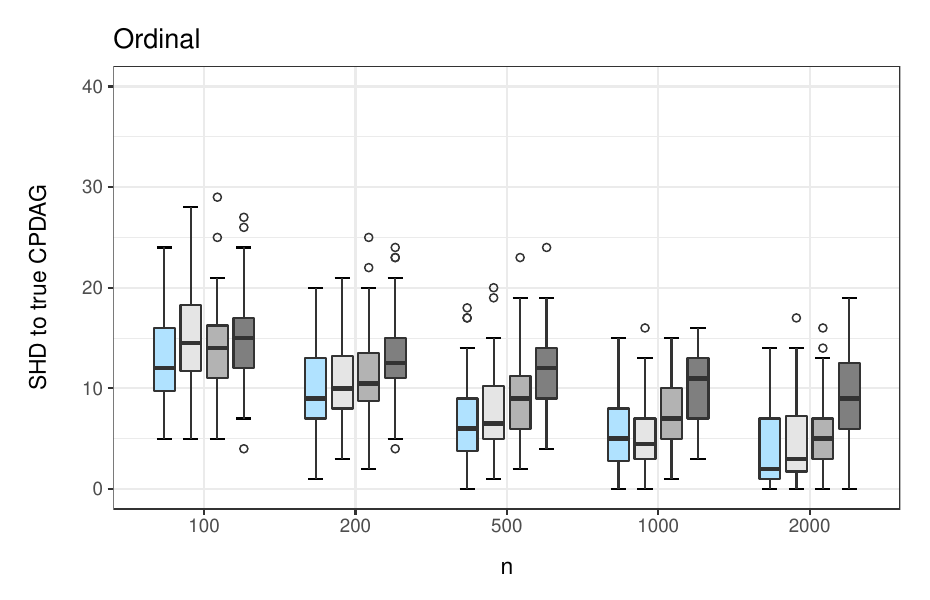}
					\\
					\includegraphics[scale=0.56]{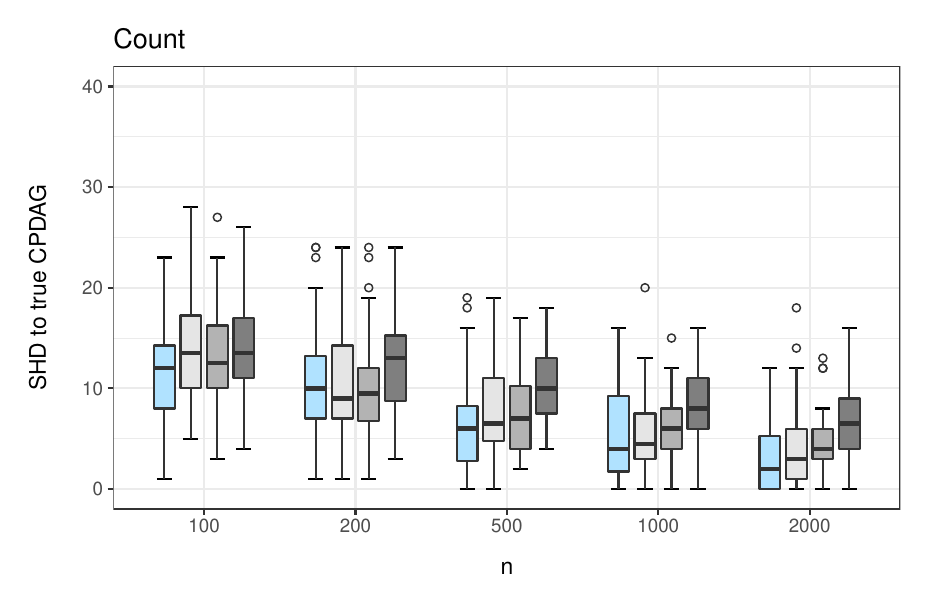}
					&
					\includegraphics[scale=0.56]{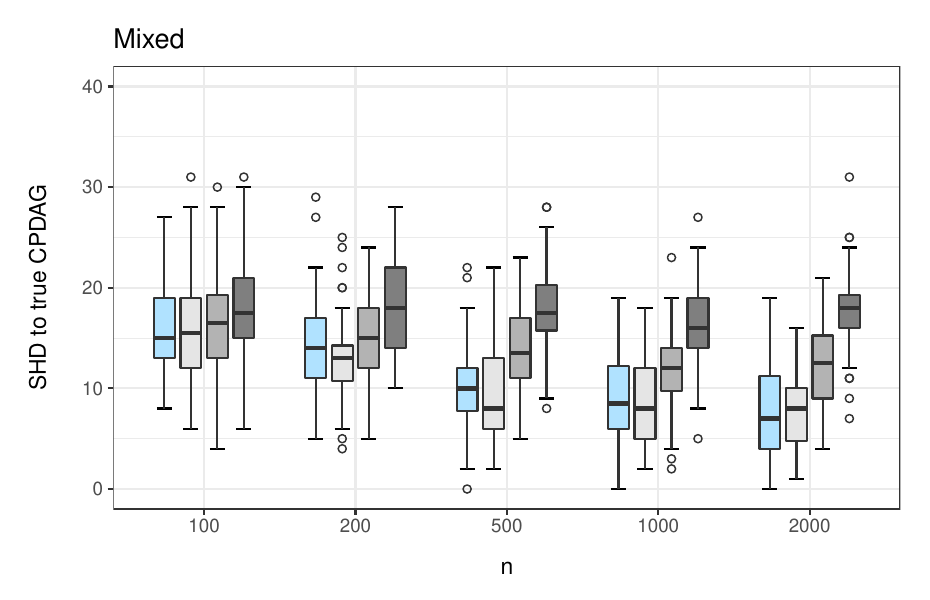}
					%\end{figure}
				\end{tabular}
				\caption{\small Simulations. Distribution (across $N=40$ simulated datasets) of the Structural Hamming Distance (SHD) between estimated and true CPDAG for type of variables \textit{Binary, Ordinal, Count, Mixed} $(b)$, sample size $n\in\{100,200,500,1000,2000\}$ $(c)$ and type of DAG structure \textit{Free}. Methods under comparison are: our Bayesian Copula DAG model (light blue) and the Copula PC method with independence tests implemented at significance level $\alpha\in\{0.01,0.0,0.10\}$ (from light to dark gray).}
				\label{fig:sim:cfr}
			\end{center}
		\end{figure}
	\end{center}
\end{landscape}

\subsection*{Comparison with Bayesian parametric strategy}

Our methodology is based on a semi-parametric strategy which
models separately the dependence parameter, corresponding to a DAG-dependent covariance matrix, and the marginal distributions of the observed variables, which are estimated using a rank-based non-parametric approach.

Alternatively, one can adopt appropriate
parametric families for modeling the various mixed types of variables, as in a generalized linear model (glm) framework.
To implement this parametric strategy, we generalize the latent Gaussian DAG-model in \eqref{eq:gaussian:DAG} to accommodate a non-zero marginal mean for the latent variables. Specifically, we assume
%We summarize below the implementation of this parametric strategy which requires a slight adaptation of our DAG-model to accommodate for a non-zero mean of each latent variable $Z_j$, as in a standard generalized linear models (glms).
%
%(this is related to the parameters of the marginal distributions that we need to estimate as well); otherwise we lose information on the location (marginal mean) of each observed variable;
%
%Implicitly, this extra term allows to include in our model (and then estimate) the parameter of the marginal distribution of each observed variable $X_j$.
%At the latent stage we assume that conditionally on a given DAG $\D$,
\ben
Z_1,\dots,Z_q \g \bmu,\bOmega, \D \sim \N_q(\bmu,\bOmega^{-1}),
\een
with $\bmu\in\Re^{q}$ and $\bOmega\in\mathcal{P}_{\D}$, the space of all s.p.d. precision matrices Markov w.r.t. DAG $\D$.
%
%To assign a prior on $(\bmu,\bOmega)$ we follow the procedure of [...].
The allied Structural Equation Model (SEM) representation of such model is given by
$\boldsymbol{\eta}+\bL^\top\bz = \boldsymbol{\varepsilon}, \boldsymbol{\varepsilon}\sim\N_q(\bzero,\bD)$,
or equivalently, in terms of node-distributions
\ben
\label{eq:SEM:bis}
Z_j = \eta_j - \bL_{\prec j \, ]}^\top\bz_{\pa_{\D}(j)} + \varepsilon_j,\quad \varepsilon_j\stackrel{\textnormal{ind}}{\sim} \N(0,\bD_{jj}),
\een
for each
$j=1,\dots,q$
with
$
\bD_{jj} = \bSigma_{j\g \pa_{\D}(j)},
\bL_{\prec j \, ]} = -\bSigma^{-1}_{\prec j \succ}\bSigma_{\prec j \, ]} ,
\eta_j = \mu_j + \bL^{\top}_{\prec j \, ]} \bmu_{\pa_{\D}(j)}.
$
Importantly, each equation in \eqref{eq:SEM:bis} now resembles the structure of a linear ``regression" model with a non-zero intercept term $\eta_j$.
A Normal-DAG-Wishart prior can be then assigned to $(\boldsymbol{\eta},\bD,\bL)$; see \citet[Supplement, Section 1]{Castelletti:Consonni:2023:bnp} for full details.
Under such prior, the posterior distribution of $(\boldsymbol{\eta},\bD,\bL)$ given independent (latent) Gaussian data $\bZ$ is still Normal-DAG-Wishart and also a marginal data distribution is available in closed-from expression.
Therefore, we can adapt the MCMC scheme of Section 4 to this more general framework and specifically with the update of $(\bD,\bL,\D)$ in Section 4.1 replaced by $(\boldsymbol{\eta}, \bD,\bL,\D)$.

Consider now the observed variables $X_1,\dots,X_q$, where each $X_j\sim F_j(\cdot)$, a suitably-specified parametric family for $X_j$, e.g. Bernoulli, Poisson, or Binomial; see also Section 4.1. %\ref{sec:simulation:study:scenarios}.
As in a glm framework, we assume that
\ben
\label{eq:link:function:glm}
\vat(X_j\g \bz_{\pa_{\D}(j)}) = h^{-1}(\eta_j - \bL_{\prec j \, ]}^\top\bz_{\pa_{\D}(j)}) 
\een
where $h^{-1}(\cdot)$ is a suitable inverse-link function and it appears that $\eta_j - \bL_{\prec j \, ]}^\top\bz_{\pa_{\D}(j)}$ plays the role of the linear predictor in the glm model for $X_j$.
Specifically, we take $h(\cdot)=\textnormal{logit}(\cdot)$
and $h(\cdot)=\log(\cdot)$
for $X_j \sim\textnormal{Bern}(\pi_j)$ and $X_j\sim\textnormal{Pois}(\lambda_j)$ respectively.
Moreover, for $X_j \sim\textnormal{Bin}(n_j,\pi_j)$ we take $h(\pi_j)=\textnormal{logit}(\pi_j)$ while fix $n_j=\max\{x_{i,j}, i=1,\dots,n\}$.
From
\eqref{eq:link:latent:observed}
we then have
$
Z_j = \Phi^{-1}\left\{F_j(X_j\g \bz_{\pa_{\D}(j)})\right\},
$
with $\Phi(\cdot)$ the standard normal c.d.f. and with $F_j$ implicitly depending on DAG parameters $(\eta_j,\bL_{\prec j \, ]})$ through \eqref{eq:link:function:glm}.
The update of $\bZ$ in Section 4.2 conditionally on the DAG parameters is then replaced by computing
$z_{i,j}=\Phi^{-1}\left\{F_j(x_{i,j}\g \bz_{\pa_{\D}(j)})\right\}$ iteratively for each $i=1,\dots,n$ and $j=1,\dots,q$.

We consider the same simulation settings as in the \textit{Balanced Scenario}, with the four different types of variables and with class of DAG structure \textit{Free}; see Section 5.1.
We compare the performance of the parametric strategy introduced above with our original method. Specifically, from the MCMC output provided by each method we first recover a CPDAG estimate and compare true and estimated graphs in terms of Structural Hamming Distance (SHD); see also Section 5.2 for details.

\begin{center}
	\begin{figure}
		\begin{center}
			\begin{tabular}{cc}
				%\textit{Sparse Scenario} \quad\quad
				%\\
				\includegraphics[scale=0.62]{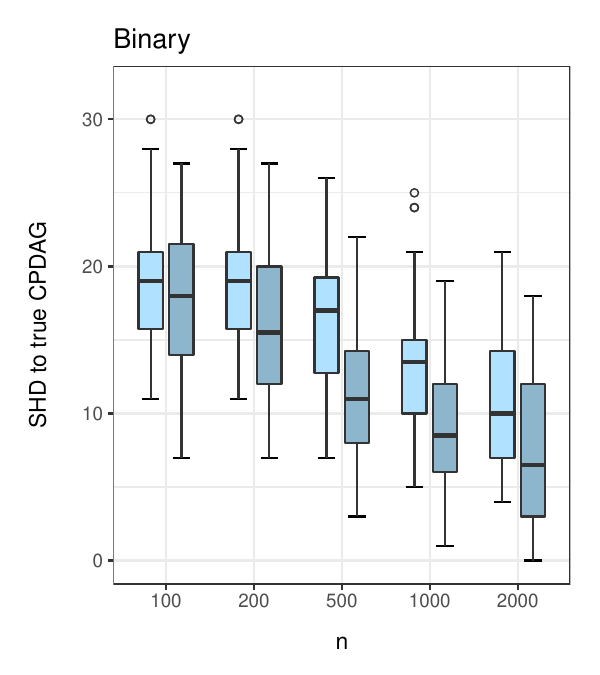}
				&
				\includegraphics[scale=0.62]{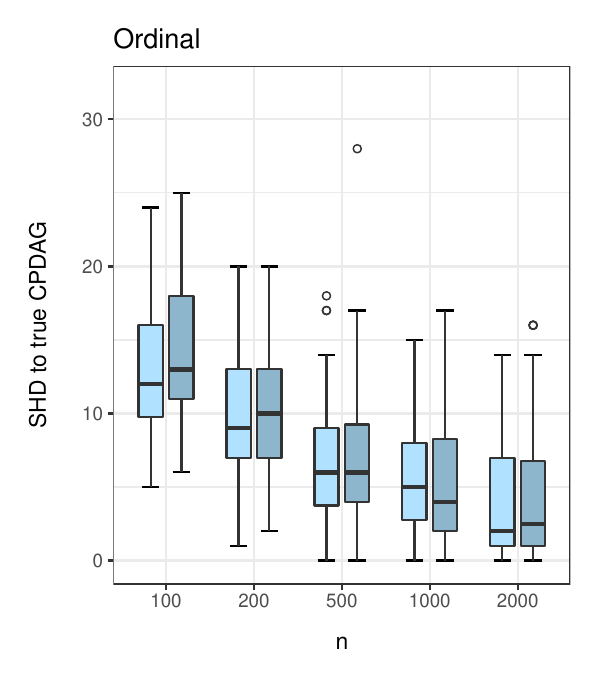}
				\\
				\includegraphics[scale=0.62]{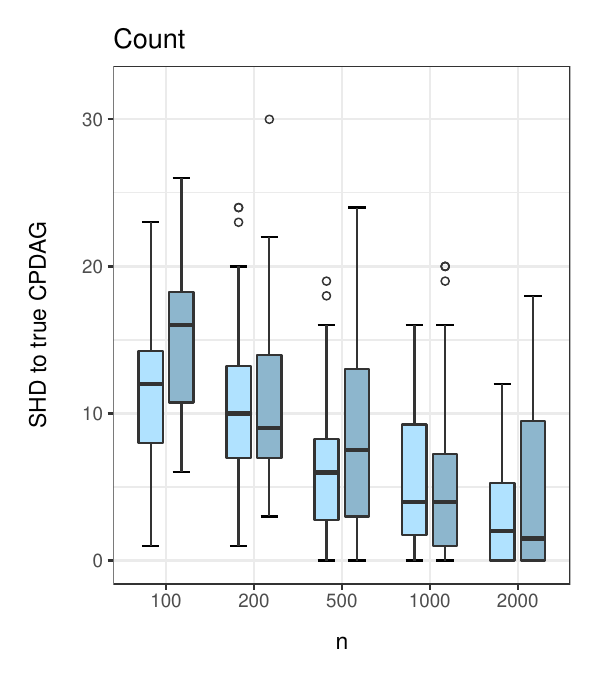}
				&
				\includegraphics[scale=0.62]{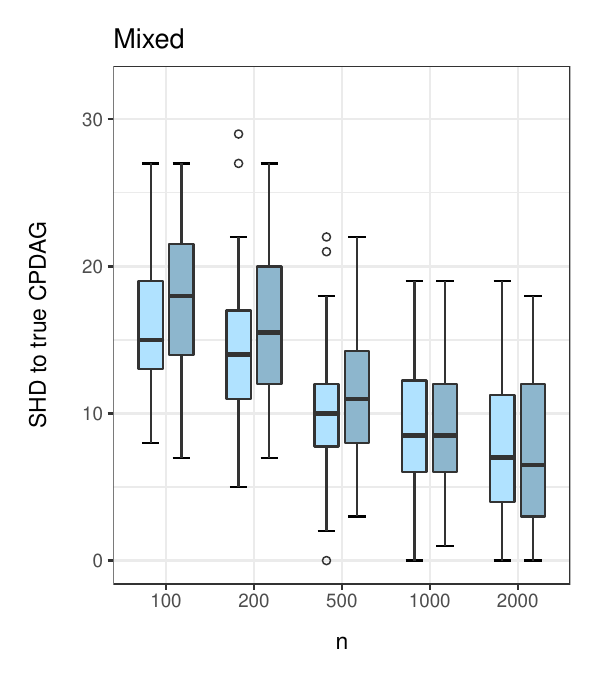}
				%\end{figure}
			\end{tabular}
			\caption{\small Simulations. Distribution (across $N=40$ simulated datasets) of the Structural Hamming Distance (SHD) between estimated and true CPDAG for type of variables \textit{Binary, Ordinal, Count, Mixed} $(b)$, sample size $n\in\{100,200,500,1000,2000\}$ $(c)$ and type of DAG structure \textit{Free}. Methods under comparison are: our original semi-parametric Bayesian Copula DAG model (light blue) and its modified version based on parametric assumptions (dark blue).}
			\label{fig:sim:cfr:parametric}
		\end{center}
	\end{figure}
\end{center}

Results are summarized in the box-plots of Figure \ref{fig:sim:cfr:parametric}, representing the distribution of SHD (across the $40$ independent simulations) obtained from our original method (light blue) and its parametric version (dark blue) under the various scenarios.
It appears that the parametric ``version" of our method outperforms our original semi-parametric model in the \textit{Binary} Scenario, while it is clearly outperformed under all the other scenarios for small-to-moderate sample sizes; however, the two approaches tend to perform similarly as the sample size $n$ increases.

\black

\subsection*{MCMC diagnostics of convergence and computational time}
\label{appendix:diagnostics}

Our methodology relies on Markov Chain Monte Carlo (MCMC) methods to approximate the posterior distribution of the parameters.
Accordingly, diagnostics of convergence of the resulting MCMC output to the target distribution should be implemented before posterior analysis.
In the following we include a few results relative to the application of our method to the well-being datas presented in Section 6.1.

\begin{center}
	\begin{figure}
		\begin{center}
			\begin{tabular}{c}
				%\textit{Sparse Scenario} \quad\quad
				%\\
				\includegraphics[scale=0.60]{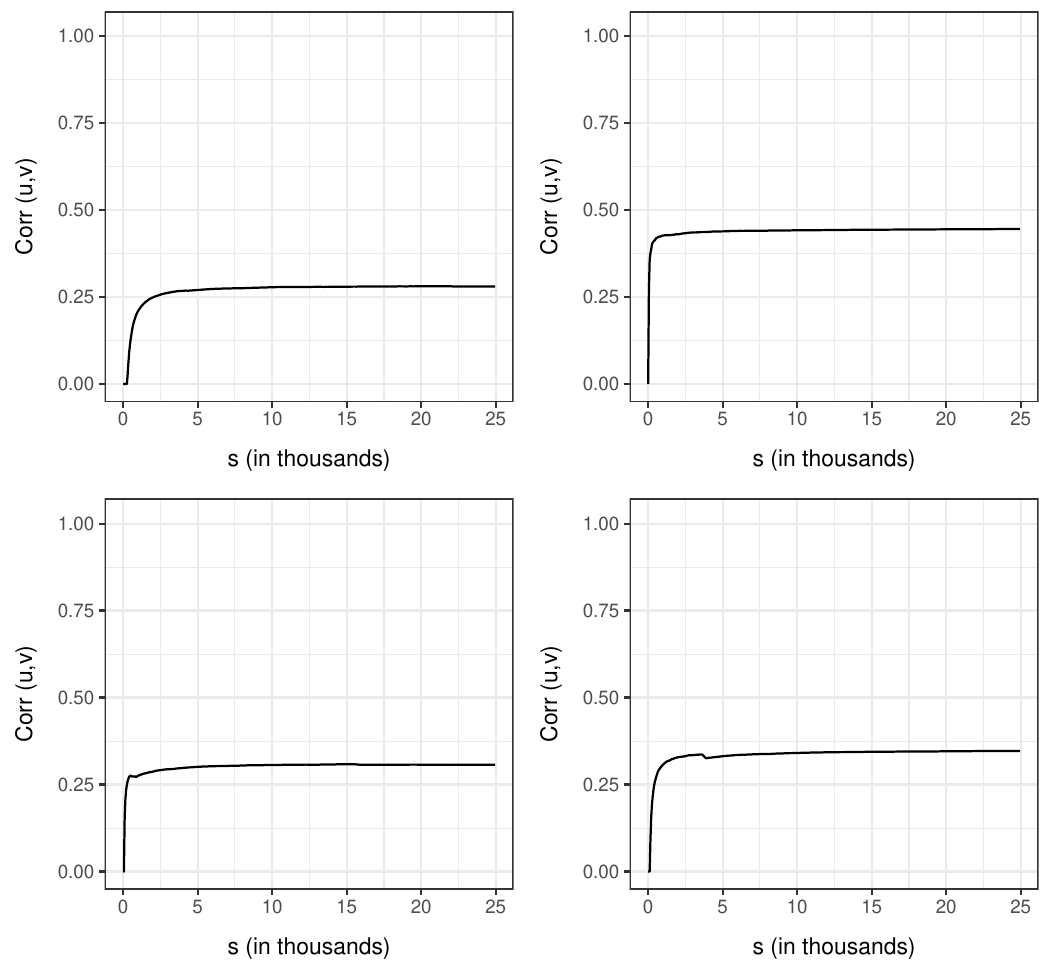}
				%\end{figure}
			\end{tabular}
			\caption{\small Well being data. Trace plots of the posterior mean of four correlation coefficients (for randomly selected variables $X_u,X_v$) estimated from the MCMC output up to iteration $s$, for $s=1,\dots,25000$.}
			\label{fig:diagnostics:1}
		\end{center}
	\end{figure}
\end{center}

As a first diagnostic tool, we monitor the behavior of the estimated posterior expectation of each correlation coefficient across iterations. Each quantity is computed at MCMC iteration $s$ using the sampled values collected up to step $s$, for $s=1,\dots,25000$.
According to the results, reported for selected variables $(X_u,X_v)$ in Figure \ref{fig:diagnostics:1}, we discard the initial $B=5000$ draws that are therefore used as a burnin period. The behavior of each traceplot suggests for each parameter an appreciable degree of convergence to the posterior mean.

\begin{center}
	\begin{figure}
		\begin{center}
			\begin{tabular}{cc}
				%\textit{Sparse Scenario} \quad\quad
				%\\
				\includegraphics[scale=0.40]{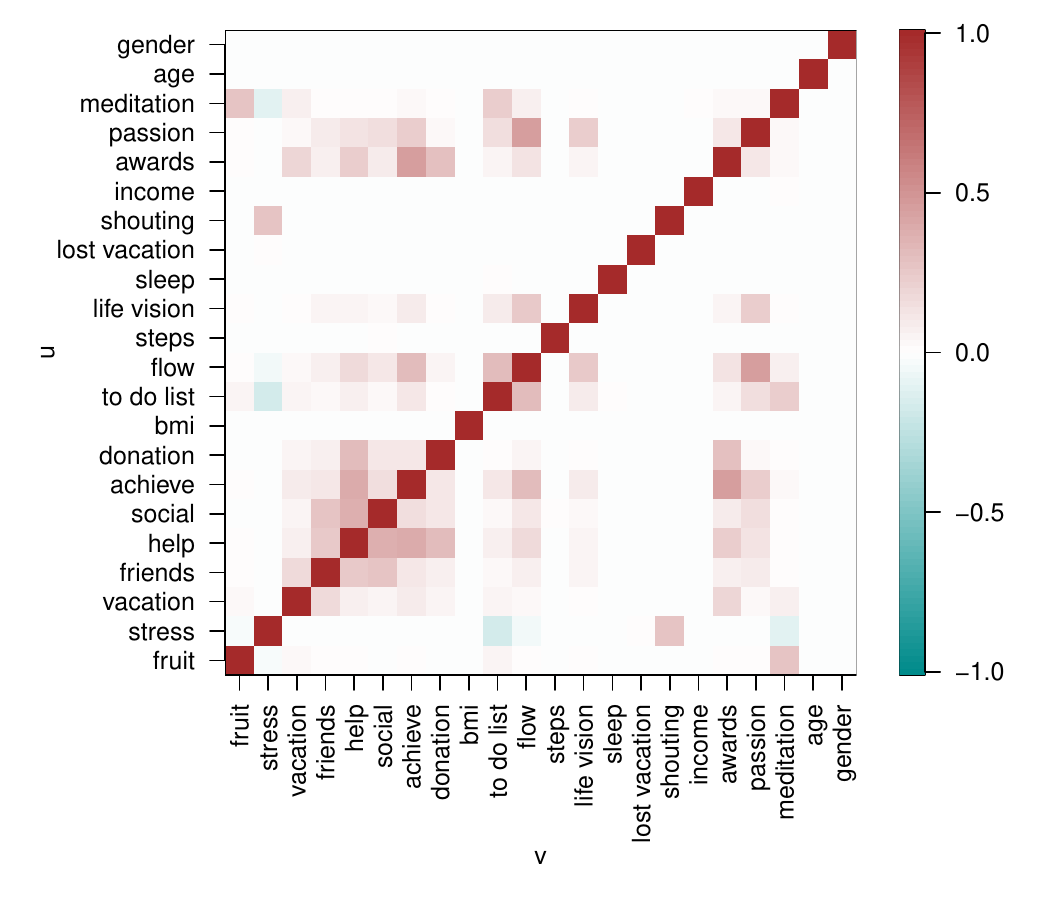}
				&
				\includegraphics[scale=0.40]{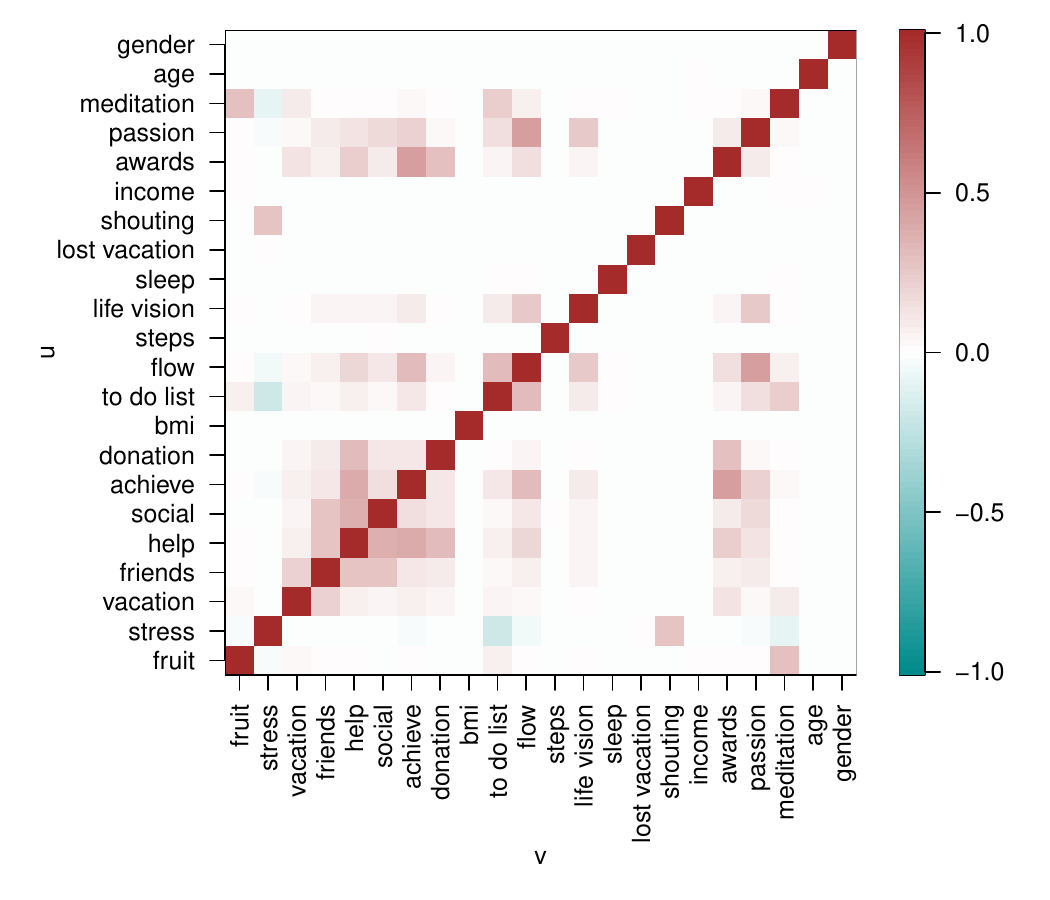}
				%\end{figure}
			\end{tabular}
			\caption{\small Well being data. Estimated correlation matrices obtained under two independent MCMC chains.}
			\label{fig:diagnostics:2}
		\end{center}
	\end{figure}
\end{center}

As a further diagnostic, we run two independent MCMC chains of length $S=25000$, again including a burnin period of $B=5000$ runs, and with randomly-chosen DAGs for the MCMC initialization.
Results in terms of estimated posterior probabilities of edge inclusion computed from the two MCMC chains are reported in the heatmpas of Figure \ref{fig:diagnostics:2} and suggest a visible agreement between the two outputs.

Finally, we investigate the computational time of our algorithm as a function of the
number of variables $q$ and sample size $n$. The following plots summarize the behavior
of the running time (averaged over $40$ replicates) per iteration, as a function of $q\in\{5,10,20,50,100\}$ for $n = 500$, and as a function of $n\in\{50,100,200,500,1000\}$ for
$q = 20$. Results were obtained on a PC Intel(R) Core(TM) i7-8550U 1,80 GHz.

\begin{center}
	\begin{figure}
		\begin{center}
			\begin{tabular}{c}
				\includegraphics[scale=0.75]{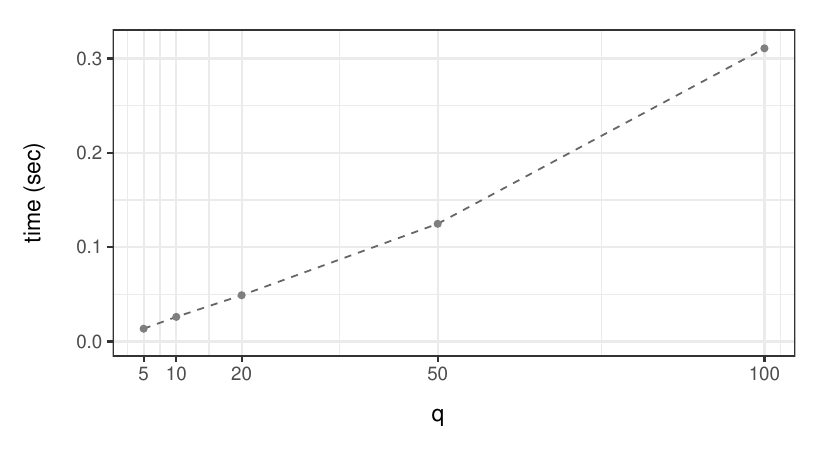}\\
				\includegraphics[scale=0.75]{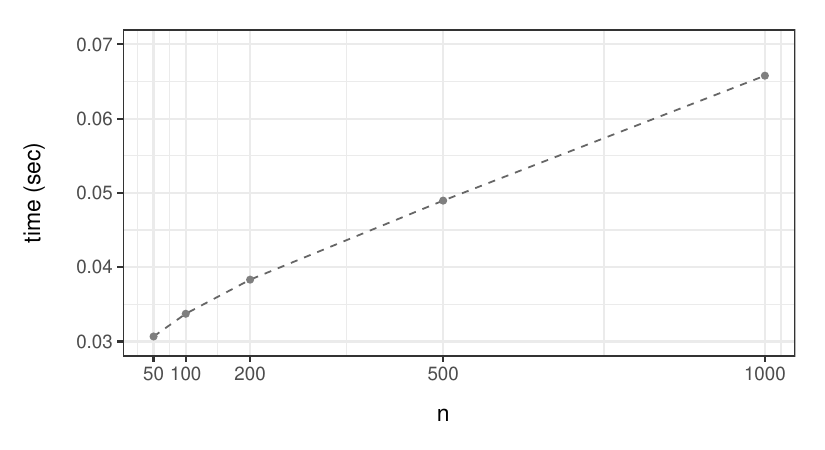}
			\end{tabular}
			\caption{\small Computational time (in seconds) per iteration, as a function of the number of variables
				$q$ for fixed $n = 500$ (upper plot) and as a function of the sample size $n$ for fixed $q = 20$ (lower plot), averaged over $40$ simulated datasets.}
			\label{fig:comp:time}
		\end{center}
	\end{figure}
\end{center}

\black

%% ITEM 9 [See the "howto.tex" file.]
%\appendix
%\renewcommand{\theequation}{A\arabic{equation}}
%\setcounter{equation}{0}
%\renewcommand{\thesection}{\Alph{subsection}}
%\setcounter{section}{0}
%\section*{Appendix}
%\section*{Appendix A}
%\section*{Appendix B}
%\vspace{\fill}\pagebreak

%% ITEM 10 [See the "howto.tex" file.]

%\begin{thebibliography}

\bibliographystyle{biometrika}
\bibliography{biblio}

%\end{thebibliography}

%% ITEM 11 [See the "howto.tex" file.]
%%%% You can put your Figures and Tables here
%%%% after the Reference Section.
%%%% BE SURE TO MARK IN THE TEXT WHERE
%%%% YOU WANT EACH FIGURE AND TABLE TO BE PLACED.
%%%% If you prefer, you can integrate your figures and tables into the text of your paper,
%%%% PROVIDED you will provide camera-ready copies of each figure.
%\vspace{\fill}\pagebreak
%\linespacing{1}

%\section*{Figures}
%
%\begin{figure}[h]
%\centerline{\includegraphics{figure01.eps}}
%\caption{Your figure caption goes here.}
%\end{figure}
%\vskip6pt

%\vspace{\fill}\pagebreak

%\section*{Tables}

%\vspace{\fill}\pagebreak

\end{document}